\documentclass[aps,pre,10pt,twocolumn,nofootinbib,showpacs]{revtex4-1}
\usepackage{graphicx}
\usepackage{amssymb}
\usepackage{amsmath}
\usepackage{amsfonts}
\usepackage{color}
\usepackage[latin1]{inputenc}
\usepackage{float}
\usepackage{comment}
\usepackage{tabularx}

\newcommand{\expct}[1]{\langle{#1}\rangle}

\begin{document}
\title{Height distributions in interface growth: The role of the averaging process}

\author{Tiago J. Oliveira}
\email{tiago@ufv.br}
\affiliation{Departamento de F\'isica, Universidade Federal de Vi\c cosa, 36570-900, Vi\c cosa, Minas Gerais, Brazil}

\begin{abstract}
Height distributions (HDs) are key quantities to uncover universality and geometry-dependence in evolving interfaces. To quantitatively characterize HDs, one uses adimensional ratios of their first central moments ($m_n$) or cumulants ($\kappa_n$), especially the skewness $S$ and kurtosis $K$, whose accurate estimate demands an averaging over all $L^d$ points of the height profile at a given time, in translation-invariant interfaces, and over $N$ independent samples. One way of doing this is by calculating $m_n(t)$ [or $\kappa_n(t)$] for each sample and then carrying out an average of them for the $N$ interfaces, with $S$ and $K$ being calculated only at the end. Another approach consists in directly calculating the ratios for each interface and, then, averaging the $N$ values. It turns out, however, that $S$ and $K$ for the growth regime HDs display strong finite-size and -time effects when estimated from these ``interface statistics'', as already observed in some previous works and clearly shown here, through extensive simulations of several discrete growth models belonging to the EW and KPZ classes on one- and two-dimensional substrates of sizes $L=const.$ and $L \sim t$. Importantly, I demonstrate that with an ``1-point statistics'', i.e., by calculating $m_n(t)$ [or $\kappa_n(t)$] once for all $N L^d$ heights together, these corrections become very weak, so that $S$ and $K$ attain values very close to the asymptotic ones already at short times and for small $L$'s. However, I find that this ``1-point'' approach fails in uncovering the universality of the HDs in the steady state regime (SSR) of systems whose average height, $\bar{h}$, is a fluctuating variable. In fact, as demonstrated here, in this regime the 1-pt height evolves as $h(t) = \bar{h}(t) + s_{\lambda} A^{1/2} L^{\alpha} \zeta + \cdots$ --- where $P(\zeta)$ is the underlying SSR HD --- and the fluctuations in $\bar{h}$ yield $S_{1pt} \sim t^{-1/2}$ and $K_{1pt} \sim t^{-1}$. Nonetheless, by analyzing $P(h-\bar{h})$, the cumulants of $P(\zeta)$ can be accurately determined. I also show that different, but universal, asymptotic values for $S$ and $K$ (related, so, to different HDs) can be found from the ``interface statistics'' in the SSR. This reveals the importance of employing the various complementary approaches to reliably determine the universality class of a given system through its different HDs.
\end{abstract}


\maketitle

\section{Introduction}
\label{secintro}

The height distribution (HD) of an evolving interface has recently been established as a measure so important to determine its universality class as the scaling properties \cite{barabasi,KrugAdv} of its roughness. In fact, the probability density functions (pdf's) for the heights of growing interfaces display a universal behavior, despite a dependence with the geometry of the system. For instance, during the transient growth regime (GR), when the correlation length $\xi$ parallel to substrate is much smaller than its lateral size $L$ ($\xi \ll L$), the height, $h$, at a given point of the interface is expected to evolve asymptotically in time, $t$, as \cite{Krug1992,Prahofer2000}
\begin{equation}
h = v_{\infty} t + s_{\lambda}(\Gamma t)^{\beta} \chi + \cdots
\label{eqAnsatzGR}
\end{equation}
where $v_{\infty}$, $s_{\lambda}$ and $\Gamma$ are system-dependent parameters, while the growth exponent $\beta$ and the pdf of the random variable $\chi$ [i.e., the underlying HD - $P(\chi)$] are universal within a given universality class, but $P(\chi)$ may change with the geometry. For example, for the one-dimensional (1D) Kardar-Parisi-Zhang (KPZ) \cite{KPZ} class, $P(\chi)$ is theoretically known to be given by Tracy-Widom (TW) distributions from a Gaussian unitary (orthogonal) ensemble [GUE (GOE)] when the interfaces are asymptotically curved (flat) \cite{Prahofer2000,Sasamoto2010,Amir,Calabrese2011}. Such behavior has been widely confirmed experimentally \cite{Takeuchi2010,Takeuchi2011} and numerically \cite{Alves11,tiago12a,Alves13,HealyCross,silvia17}, and recently generalized for circular KPZ interfaces ingrowing \cite{Fukai2017,Ismael18} or evolving out of plane \cite{Ismael19}, as well as to half-space KPZ systems \cite{Gueudre,*Borodin16,*Borodin18,*Krajenbrink}. Universal, but geometry-dependent GR HDs have been also numerically found for the 2D KPZ class \cite{healy12,tiago13,healy13,Ismael14}, as well as for the Villain-Lai-Das Sarma (VLDS) \cite{Villain,LDS} class in both 1D and 2D substrates \cite{Ismael16a}. More recently, a geometry effect was found also in the variances of the GR HDs for the linear classes by Edwards-Wilkinson (EW) \cite{EW} and Mullins-Herring (MH) \cite{Mullins,*Herring1951}, whose pdf's are Gaussian \cite{Ismael19b}.

These advances have motivated the study of GR HDs in several experimental systems. For instance, in the 1D case, they have been investigated for interfaces of paper burning fronts \cite{Miettinen2005}, of turbulent phases in liquid crystal films \cite{Takeuchi2010,Takeuchi2011} and of colloidal particles deposited at the edges of evaporating drops \cite{Yunker2013a}. In two-dimensions, they have been analyzed in vapor deposition of CdTe films on different substrates and temperatures \cite{Almeida14,Almeida15,Almeida17} and of oligomer films on Si \cite{healy14exp}, and also in electrodeposition of NiW on polished steel \cite{Rodolfo17} and of oxide films of Cu$_2$O on $n-$Si(100) and Ni/$n-$Si(100) \cite{Yuri15}. With exception of this last study, in all others evidence of KPZ scaling were found. One important lesson coming from these works is that the comparison of HDs --- as well as of other universal distributions such as of local roughness and extreme heights (see, e.g., Ref. \cite{Ismael16b} for a recent survey of literature) --- is complementary to the traditional analysis of exponents from the dynamic scaling of the surface roughness and other correlation functions. 

From a more theoretical side, GR HDs have been employed in the study of the upper critical dimension, $d_u$, of the KPZ class, strongly indicating that, if it exists, $d_u>6$ \cite{Alves14,HHTake2015}. Moreover, the analysis of HDs has proven useful in unveiling the origins of the strong corrections to scaling observed in ballistic-like models \cite{Alves14BD}. 

Furthermore, universal HDs have been also observed in the steady state regime (SSR), when $\xi \sim L$, where they might differ from the GR HDs. For example, the SSR height fluctuations (about the mean height) for periodic 1D KPZ interfaces are Gaussian distributed \cite{barabasi}. The same occurs in the EW and MH classes \cite{barabasi,KrugAdv}, whereas numerical studies for the 2D KPZ class \cite{Chin,Marinari,Fabio2004kpz,tiago07Rug} and for the VLDS class \cite{Fabio2004vlds,tiago07Rug,Singha16} reveal non-Gaussian SSR HDs in these cases.

To quantitatively compare HDs, and pdf's in general, one analyzes adimensional ratios of their first $n$th cumulants ($\kappa_n$) [or \textit{central} moments ($m_n$)], focusing mainly on the skewness $S_{\kappa} = \kappa_3/\kappa_2^{3/2}$ [or $S_m = m_3/m_2^{3/2}$] and (excess) kurtosis $K_{\kappa} = \kappa_4/\kappa_2^{2}$ [or $K_m = m_4/m_2^{2}-3$], which respectively quantify the asymmetry of the distribution and the weight of their tails when compared with a Gaussian (for which $S=K=0$). In flat and in isotropic radial growth, the interfaces are translation- and rotation-invariant, respectively, so that all of their sites are statistically equivalent. Thereby, if one has $N$ of such interfaces at hand, each one with lateral size $L$ and $d$ spatial dimensions, the ratios $S$ and $K$ (and any other) might be estimated in three main ways:
\begin{itemize}
 \item[$(A)$] We can perform a ``1-interface statistics'', by calculating the cumulants $(\kappa_{n})_i$ and, then, $(S_{\kappa})_i$ and $(K_{\kappa})_i$ for each interface $i$ at time $t$. Finally, an average over the different interfaces (denoted here by $\expct{\cdots}$) is carried out to give $S_{\kappa}^{(A)}=\expct{(S_{\kappa})_i}$ and $K_{\kappa}^{(A)}=\expct{(K_{\kappa})_i}$. 
  
 \item[$(B)$] A similar, but different ``multi-interface statistics'' can be obtained by evaluating $\kappa_n=\expct{(\kappa_n)_i}$ and, then, using these final averages to calculate the ratios $S_{\kappa}^{(B)}$ and $K_{\kappa}^{(B)}$ at a given time $t$.
 
 \item[$(C)$] We may also perform a ``1-point statistics'', by calculating the \textit{raw} moments $\overline{h^n}$ considering all the $N L^d$ available substrate sites (for the $N$ samples) together, at a given time $t$, and only at the end estimating the cumulants and then $S_{\kappa}^{(C)}$ and $K_{\kappa}^{(C)}$. 
\end{itemize}

\noindent The very same procedures can be employed for the central moments, rather than for the cumulants, to estimate $S_{m}^{(j)}$ and $K_{m}^{(j)}$, with $j=A$, $B$ or $C$. Note that, even though these methods are identical if a single sample ($N=1$) is analyzed, the ensemble average breaks their equality. At first sight, when $N>1$, approach $(C)$ could be expected to be better than the others in general, once it does not involve averages of sample-averaged quantities. It turns out that in some systems it is not always possible to warrant the statistical equivalence of different samples --- for example, AFM images taken at different regions of a thin film surface might have appreciable differences in their average height $\bar{h}_i$ ---, so that their heights can not be treated together as in method $(C)$. So, given the relevance of HDs, it is important to investigate in detail the outcomes from $(A)$, $(B)$ and $(C)$, to determine in which situations each approach may work better.

In this work, I present a deep comparison of these averaging processes and quantities (moments \textit{vs.} cumulants), based on some analytical approaches and extensive Monte Carlo simulations of several discrete models belonging to the KPZ and EW classes, on both 1D and 2D substrates, considering flat and curved geometries. Substantially, I demonstrate that methods $(A)$, $(B)$ and $(C)$ may yield quite different results for the skewness and kurtosis of a given ensemble of finite interfaces. For example, it is shown here that while approaches $(A)$ and $(B)$ may introduce strong finite-size and -time effects in $S$ and $K$ for the GR HDs, in approach $(C)$ such effects are very weak. Conversely, method $(C)$ may fail in reveling the universality of the SSR HDs, while approaches $(A)$ and $(B)$ work fine in this regime, where they may yield different ratios and, so, different height fluctuations. Motivated by the ``KPZ ansatz'' in Eq. \ref{eqAnsatzGR} for the GR HDs, I also introduce here its counterpart for the SSR, from which the underlying KPZ HDs about the mean are characterized.

The rest of the paper is organized as follows. In Sec. \ref{secModels}, I define the investigated models and simulation methods, as well as the quantities analyzed here. Numerical results for the KPZ HDs in flat and curved geometries are presented in Secs. \ref{secKPZflat} and \ref{secKPZcurved}, respectively. Section \ref{secDisc} brings a detailed comparison of the different averaging procedures, based mainly on analytical calculations, which explain their different outcomes. An ansatz for the 1-pt height in the SSR is also introduced in this section. A summary of the main results and some concluding remarks are given in Sec. \ref{secConc}.

\section{Models and quantities of interest}
\label{secModels}

I investigate three models belonging to the KPZ class: the restricted solid-on-solid (RSOS) model \cite{KK}, the ballistic deposition (BD) model \cite{barabasi} and the etching model by Mello \textit{et al.} \cite{Mello01}; and also the EW model by Family \cite{Family}. All these models were simulated on 1D arrays of $L$ sites ($d=1$) and on square lattices of lateral size $L$ ($d=2$), with periodic boundary conditions (PBC). In all cases, particles are sequentially released vertically toward the horizontal substrate, at randomly chosen positions, say $i$, and aggregate there following the rules: \textit{RSOS}: $h_i\rightarrow h_i + 1$ if $(h_i - h_j) < 1$ $\forall$ nearest neighbors (NNs) $j$; otherwise, the deposition attempt is rejected. \textit{BD}: $h_i\rightarrow max[h_i + 1,h_j]$ $\forall$ NNs $j$. \textit{Etching}: $h_j\rightarrow max[h_i,h_j]$ $\forall$ NNs $j$ and, then, $h_i\rightarrow h_i + 1$. \textit{Family}: $h_i\rightarrow h_i + 1$ if $h_i \leq h_j$  $\forall$ NNs $j$; otherwise, the NN $j$ with minimal height is taken and $h_{j}\rightarrow h_{j} + 1$, with a random draw resolving possible ties.

For all models, substrates of fixed sizes $L$ ranging from $L=32$ to $8192$ were considered, with a number $N$ of independent samples such that $N L^d \geqslant 10^9$. Results for the 1D KPZ models simulated on expanding substrates will be also discussed here. In this case, the deposition rules just described are stochastically mixed with duplications of randomly chosen columns of the substrate, to make its size evolves as $\expct{L} = L_0 + \omega t$. For more details on the implementation of these processes, I invite the reader to see the Refs. \cite{Ismael14,Ismael16a,Ismael19b}. Here, I will consider the parameters $L_0=\omega$, with $\omega$ ranging from $\omega=1$ to $20$. A large number $N$ of samples is simulated, yielding a statistics comparable to that of the flat (fixed $L$) case.

In all cases, one starts the growth with a flat initial condition, i.e., $h_i(t=0)=0$ $\forall$ substrate site $i$. After each deposition attempt (or column duplication in expanding systems) the time is updated by $t \rightarrow t + \Delta t$, with $\Delta t = 1/(L^d + \omega d)$, where $\omega = 0$ in the case of fixed size substrates.

Note that if $h_{ij}$, with $j = 1,\ldots,L^d$, represents the height profile of the interface $i$, at a given time $t$, so its $n$th \textit{raw} moment is
\begin{equation}
 (\overline{h^n})_i = \frac{1}{L^d} \sum_{j=1}^{L^d} h_{ij}^n,
\end{equation}
its $n$th \textit{central} moment reads
\begin{equation}
 (m_n)_i = \frac{1}{L^d} \sum_{j=1}^{L^d} [h_{ij} - \bar{h}_i]^n,
\end{equation}
and its first cumulants are given by
\begin{subequations}
\begin{eqnarray}
 (\kappa_1)_i &=& \bar{h}_i, \\
 (\kappa_2)_i &=& (\overline{h^2})_i - \bar{h}_i^2 = (m_2)_i \equiv w_2, \\
 (\kappa_3)_i &=& (\overline{h^3})_i - 3 \bar{h}_i (\overline{h^2})_i + 2 \bar{h}_i^3 = (m_3)_i, \\
 (\kappa_4)_i &=& (\overline{h^4})_i - 4 \bar{h}_i (\overline{h^3})_i - 3 (\overline{h^2})_i^2 \\ \nonumber
 &+& 12 \bar{h}_i^2 (\overline{h^2})_i - 6 \bar{h}_i^4 = (m_4)_i - 3 (m_2)_i^2,
\end{eqnarray}
\label{eqCumul}
\end{subequations}
where $w_2$ is the squared roughness of interface $i$. Hence, $(S_{\kappa})_i = (\kappa_3)_i/(\kappa_2)_i^{3/2} = (m_3)_i/(m_2)_i^{3/2} = (S_m)_i$ and $(K_{\kappa})_i = (\kappa_4)_i/(\kappa_2)_i^2 = (m_4)_i/(m_2)_i^2 - 3 = (K_m)_i$, meaning that $\expct{(S_{\kappa})_i}=\expct{(S_m)_i}$ and $\expct{(K_{\kappa})_i}=\expct{(K_m)_i}$, where $\expct{X_i} = \frac{1}{N} \sum_{i=1}^N X_i$. Therefore, $S_{\kappa}^{(A)}=S_m^{(A)}=S^{(A)}$ and $K_{\kappa}^{(A)}=K_m^{(A)}=K^{(A)}$. 

In procedure $(B)$, one deals with $\kappa_n = \expct{(\kappa_n)_i}$ and $m_n = \expct{(m_n)_i}$. From Eqs. \ref{eqCumul}, one has $\kappa_2 = m_2 = \expct{w_2}$ and $\kappa_3 = m_3$, so that $S_{\kappa}^{(B)}=S_m^{(B)}=S^{(B)}$. However, for the kurtosis
\begin{equation}
 K_{\kappa}^{(B)} = \frac{\kappa_4}{\kappa_2^2} = \frac{m_4}{m_2^2} - 3\frac{\expct{w_2^2}}{\expct{w_2}^2} = K_m^{(B)} - 3 R_2,
 \label{eqKAKB}
\end{equation}
where
\begin{equation}
 R_n \equiv \frac{\expct{w_2^n}-\expct{w_2}^n}{\expct{w_2}^n},
 \label{eqCoefVar}
\end{equation}
with $R_2$ being the squared coefficient of variation of the width distribution $P_{w_2}(w_2)$. Since $w_2$ is a fluctuating variable, one has $R_2 > 0$ and thus $K_{\kappa}^{(B)} < K_m^{(B)}$.

The raw moments for the set of all the $N L^d$ heights, used in approach $(C)$, are simply $\overline{h^n} = \expct{(\overline{h^n})_i}$ and the related cumulants (let us denote them as $\kappa_n^*$) follow from expressions analogous to those in Eqs. \ref{eqCumul}, but without the index $i$. Thereby, similarly to case $(A)$, one has $S_{\kappa}^{(C)}=S_m^{(C)}=S^{(C)}$ and $K_{\kappa}^{(C)}=K_m^{(C)}=K^{(C)}$.

\section{Results for KPZ models on flat (fixed size) substrates}
\label{secKPZflat}

\begin{figure}[!b]
	\includegraphics[width=4.25cm]{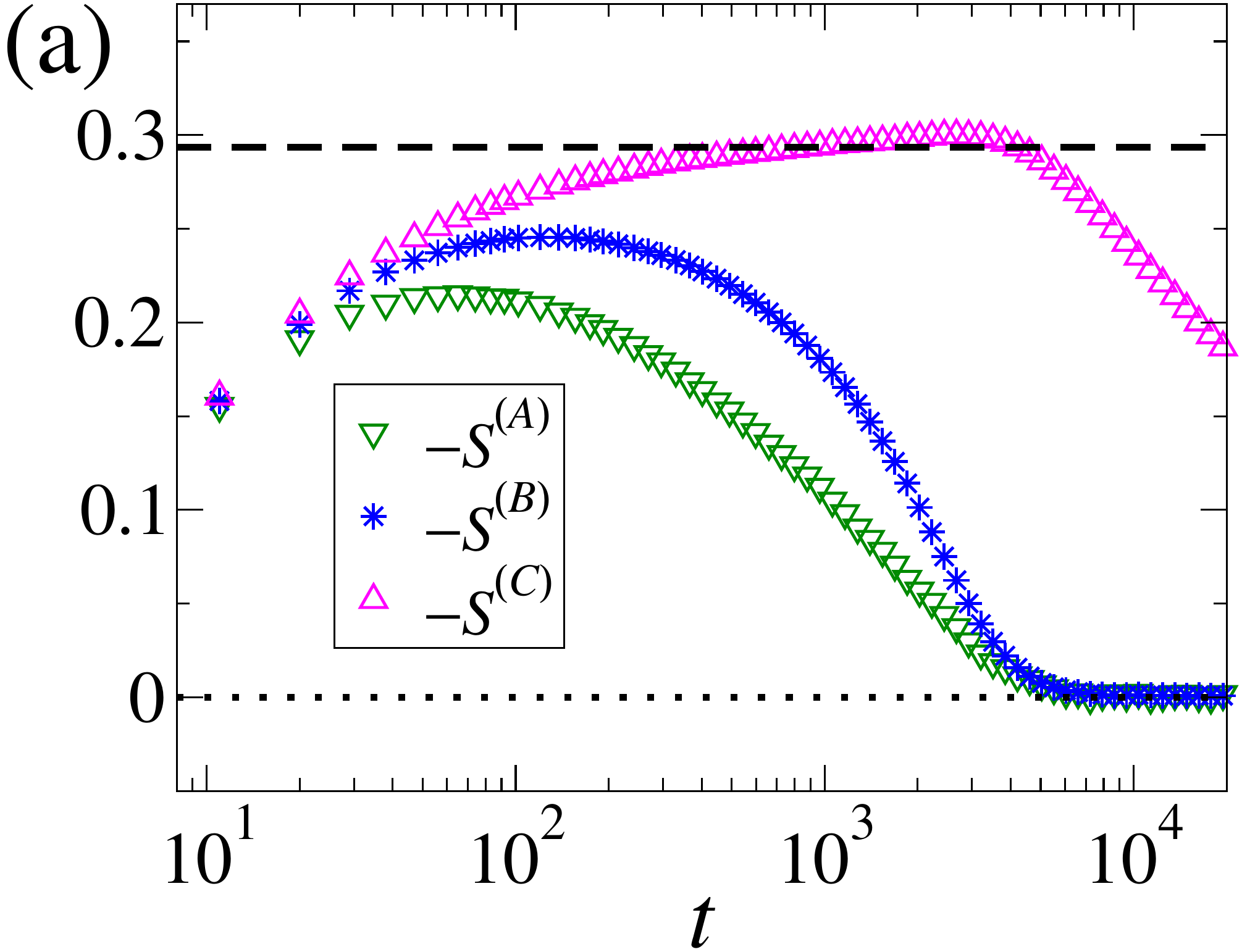}
	\includegraphics[width=4.25cm]{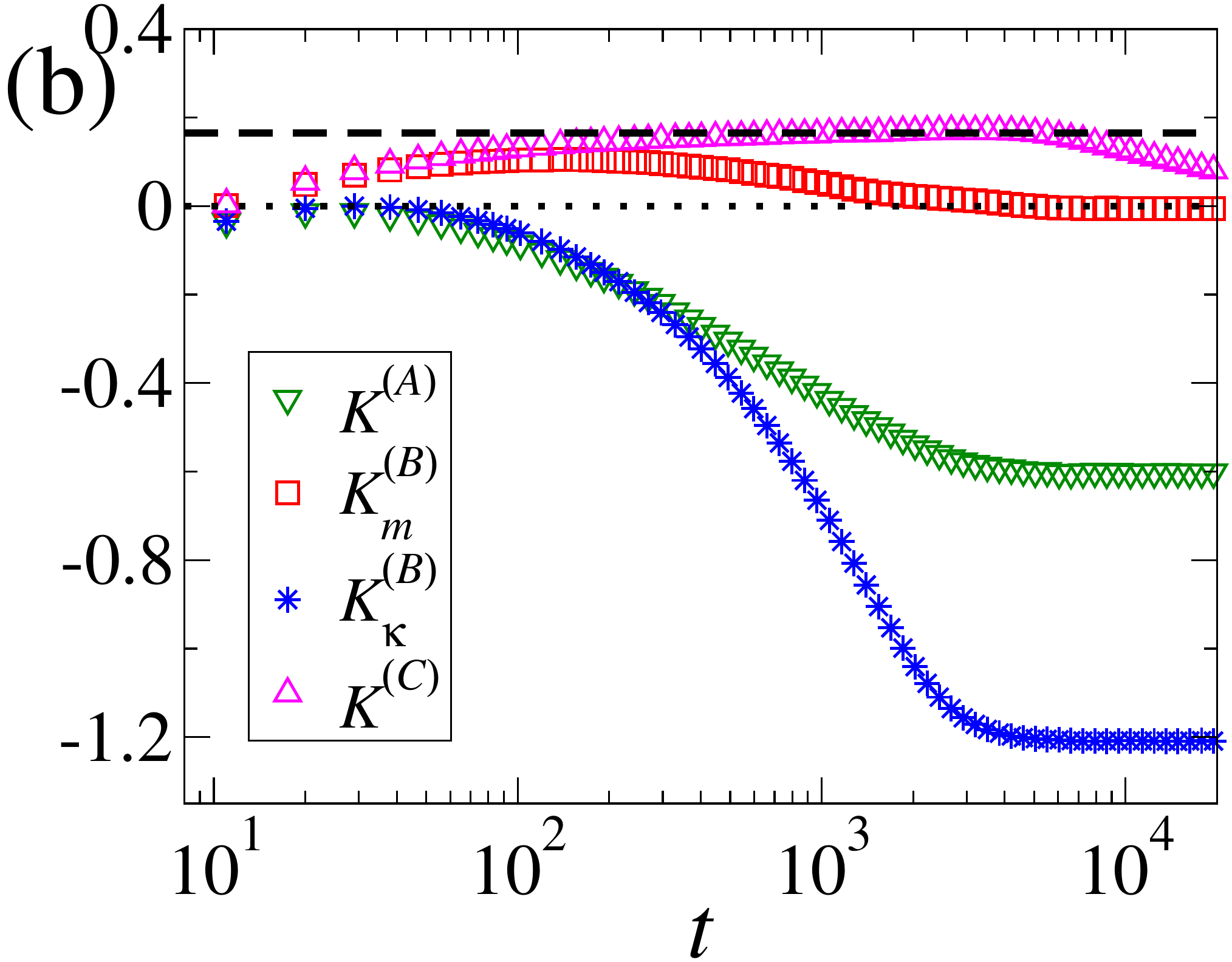}
	\caption{Temporal evolution of the (a) skewness $-S$ and (b) kurtosis $K$, for the RSOS model on a 1D substrate of size $L=512$. Data for approaches $(A)$, $(B)$ and $(C)$ are shown, as indicated. The dashed (dotted) horizontal lines represent the ratios for the TW-GOE (Gaussian) distribution.}
	\label{fig1}
\end{figure}

Figures \ref{fig1}(a) and \ref{fig1}(b) respectively show $-S$ and $K$ versus time, comparing results for the three methods [$(A)$, $(B)$ and $(C)$] defined above, for the 1D RSOS model with $L=512$. In such plots, one clearly sees that $S^{(A)}(t) \neq S^{(B)}(t) \neq S^{(C)}(t)$ and $K^{(A)}(t) \neq K_{m}^{(B)}(t) \neq K_{\kappa}^{(B)}(t) \neq K^{(C)}(t)$. Hence, three [four] different results can be obtained for $S(t)$ [$K(t)$], for the same set of interfaces, depending on the way this ratio is calculated. The same thing is found for the other KPZ models, in both 1D and 2D flat substrates, as demonstrated below. The different characteristics of each approach will be discussed separately for each regime (GR and SSR) in what follows.

\subsection{KPZ HDs for the growth regime}

\begin{figure}[!t]
	\includegraphics[width=4.25cm]{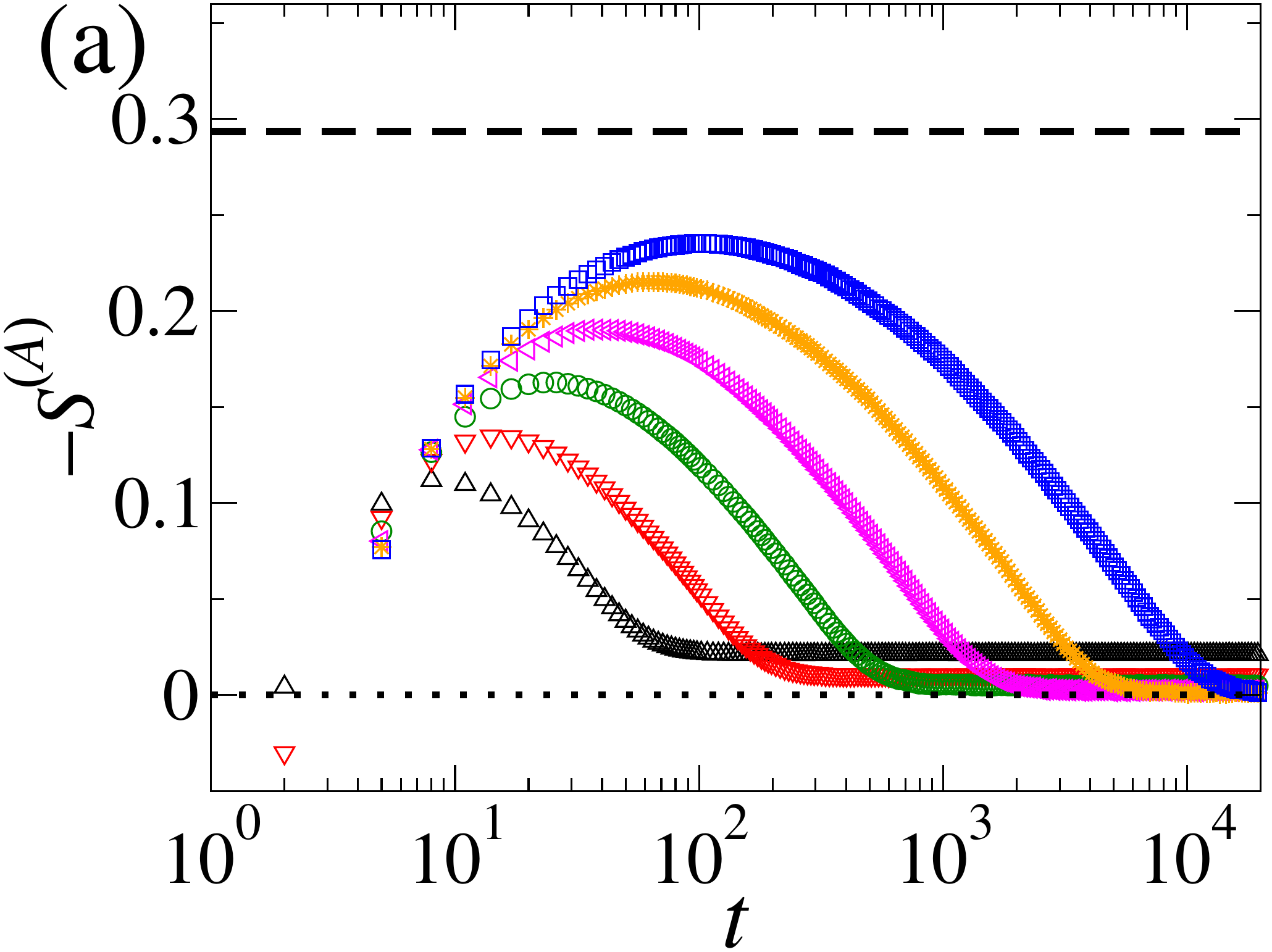}
	\includegraphics[width=4.25cm]{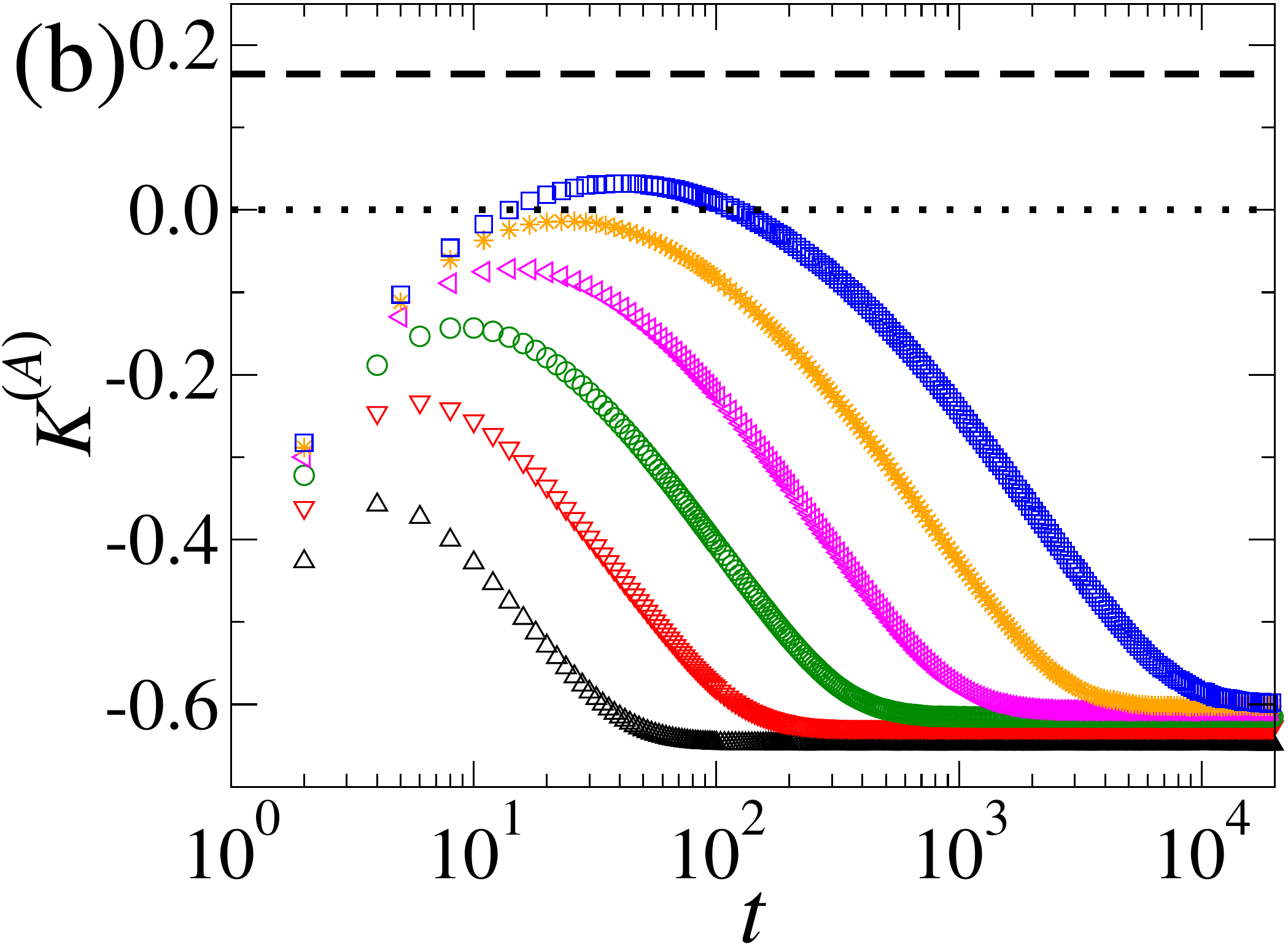}
	\includegraphics[width=4.25cm]{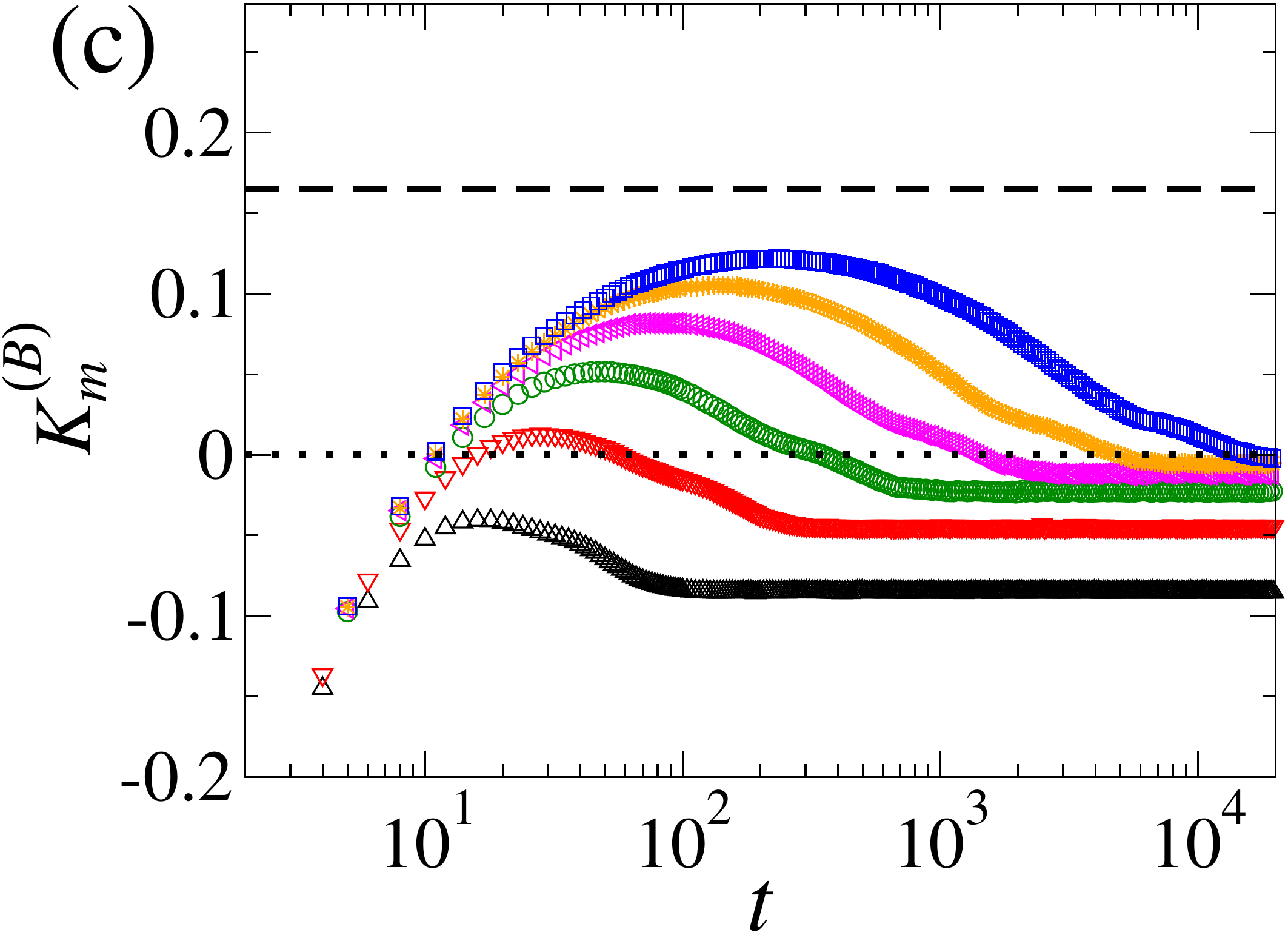}
	\includegraphics[width=4.25cm]{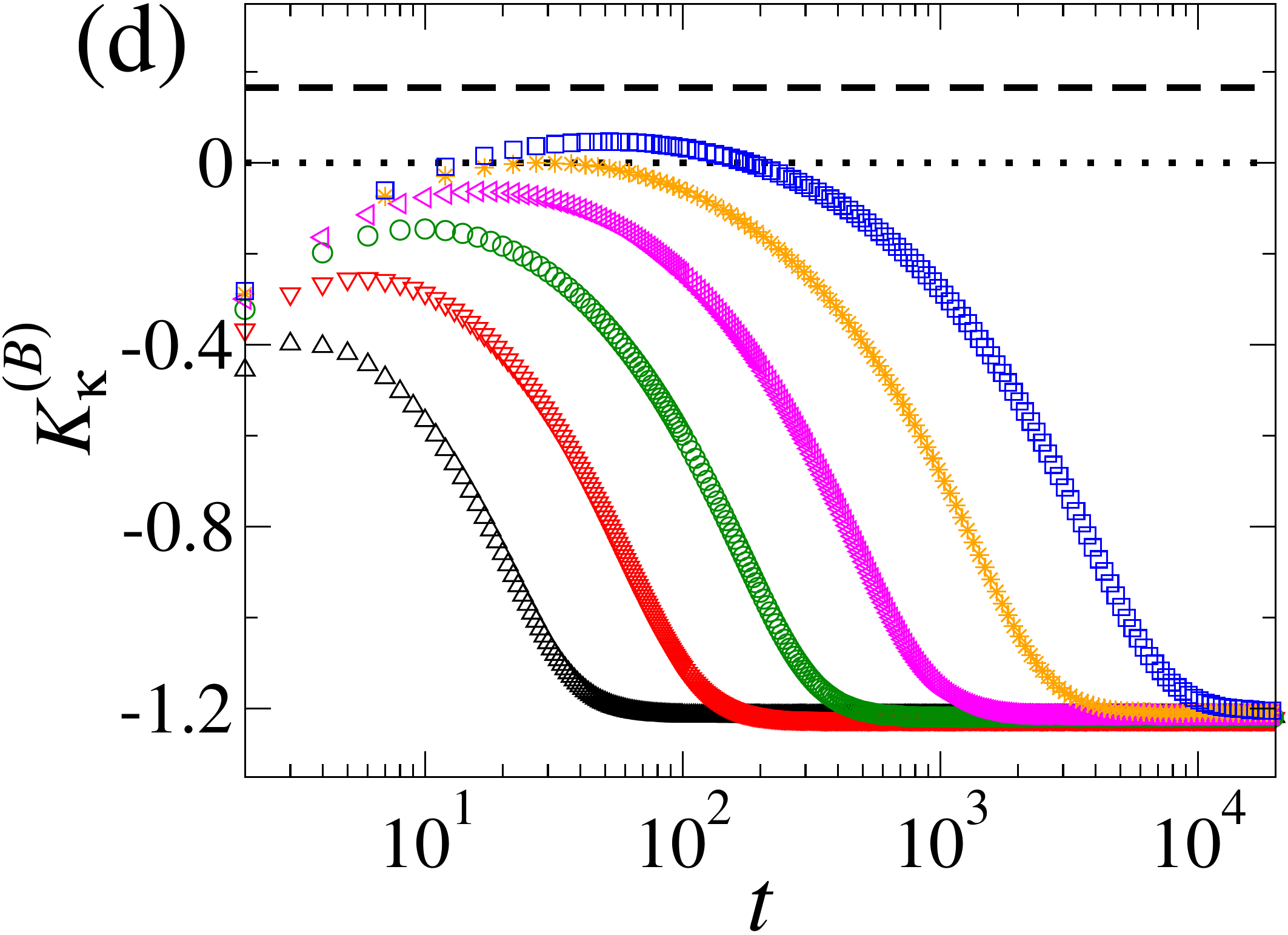}
	\includegraphics[width=4.25cm]{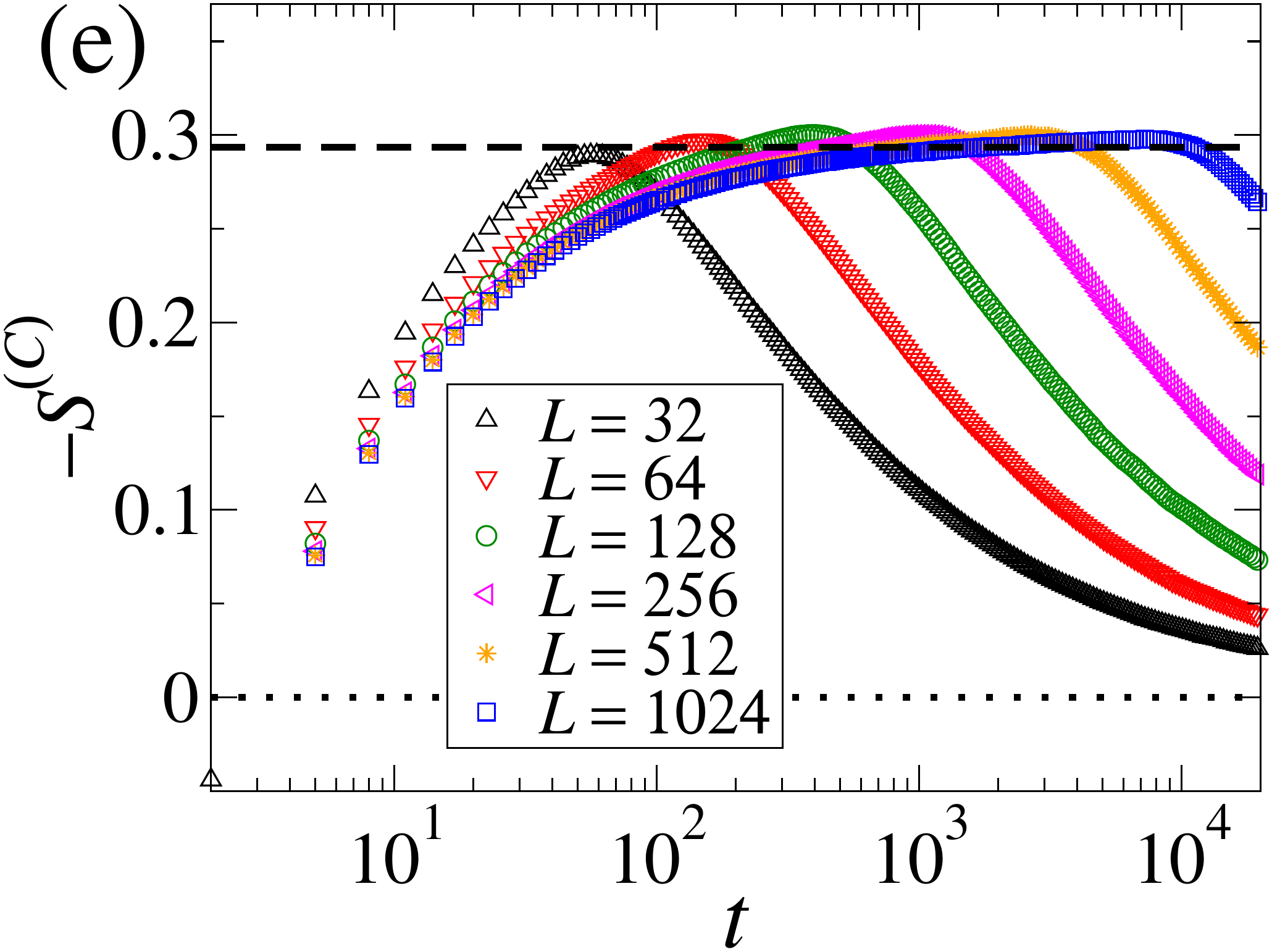}
	\includegraphics[width=4.25cm]{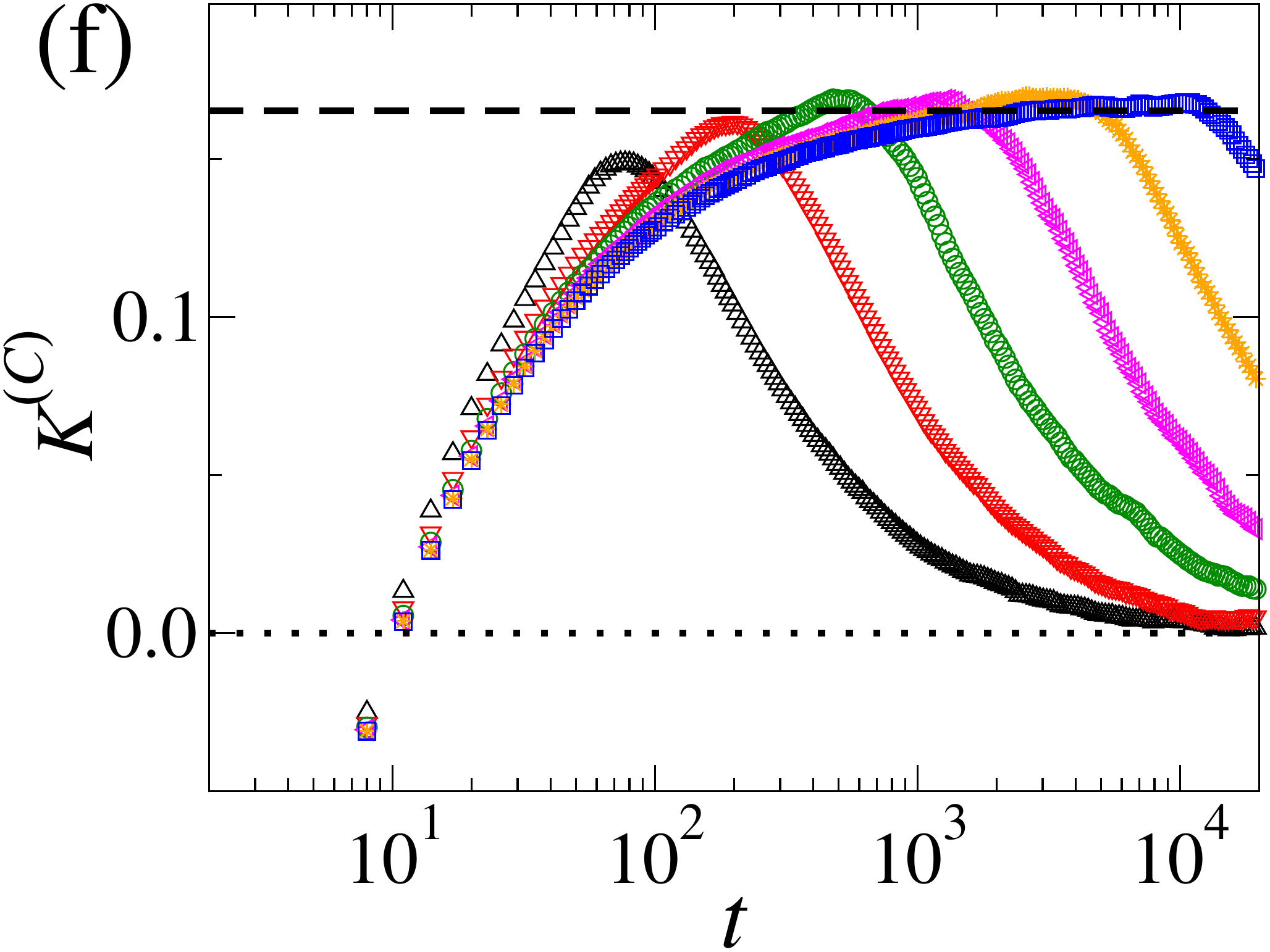}
	\caption{(a)-(f) Temporal evolution of the skewness $-S^{(x)}$ and kurtosis $K^{(x)}$, with $x=A,B,C$, for the RSOS model on 1D substrates of several sizes, as indicated in panel (e). The dashed (dotted) horizontal lines represent the ratios for the TW-GOE (Gaussian) distribution.}
	\label{fig2}
\end{figure}

Let us focus initially on a comparison of the outcomes from the three methods during the transient GR. In this case, the asymptotic HDs for the 1D KPZ models are expected to be given by the TW-GOE distribution, for which $S_{GOE}=0.2935$ and $K_{GOE}=0.1652$ \cite{Prahofer2000}, but no evidence of this is seen in Fig. \ref{fig1} for the methods $(A)$ and $(B)$. Actually, there are maxima in $-S$ and $K$ curves (which I will denote by $\tilde{S}$ and $\tilde{K}$) for these approaches, indicating that at short times both ratios start increasing toward their asymptotic values, but, due to the small $L$ analyzed, the system crosses over to the SSR before reaching there. In fact, to confirm that the asymptotic GR HD for the 1D RSOS model is the TW-GOE distribution, extensive simulations for a very large substrate size ($L=1048576$) and very long times (up to $t=500000$) had to be performed in Ref. \cite{tiago12a}, where procedure $(B)$ was employed for the cumulants. Figures \ref{fig2}a and \ref{fig2}b respectively show the temporal variation of $-S^{(A)}$ and $K^{(A)}$ for the 1D RSOS model, comparing curves for several $L$'s, and one sees that $\tilde{S}$ and $\tilde{K}$ do indeed approximate $S_{GOE}$ and $K_{GOE}$ as $L$ increases. The same thing is observed in Figs. \ref{fig2}c and \ref{fig2}d for the kurtoses from procedure $(B)$ [the behavior of $-S^{(B)}$ is analogous to that in Fig. \ref{fig2}a for $-S^{(A)}$]. These features are expected and have already been observed by Shim \& Landau \cite{LandauShim} for the method $(A)$.

The important finding here is that for method $(C)$ both $-S^{(C)}$ and $K^{(C)}$ present approximated plateaus quite close to the expected KPZ values in Fig. \ref{fig1}, demonstrating that such procedure yields results for the GR KPZ HDs with very weak finite-size and -time effects. This is indeed confirmed in Figs. \ref{fig2}e and \ref{fig2}f, which present the temporal variation of $-S^{(C)}$ and $K^{(C)}$ for the 1D RSOS model with several $L$'s, where one sees that already for $L=64$ they have a maximum quite close to the TW-GOE values. Moreover, as $L$ increases these maxima tend to give place to plateaus, since the duration of the GR increases with $L$.

\begin{figure}[!t]
	\includegraphics[width=4.25cm]{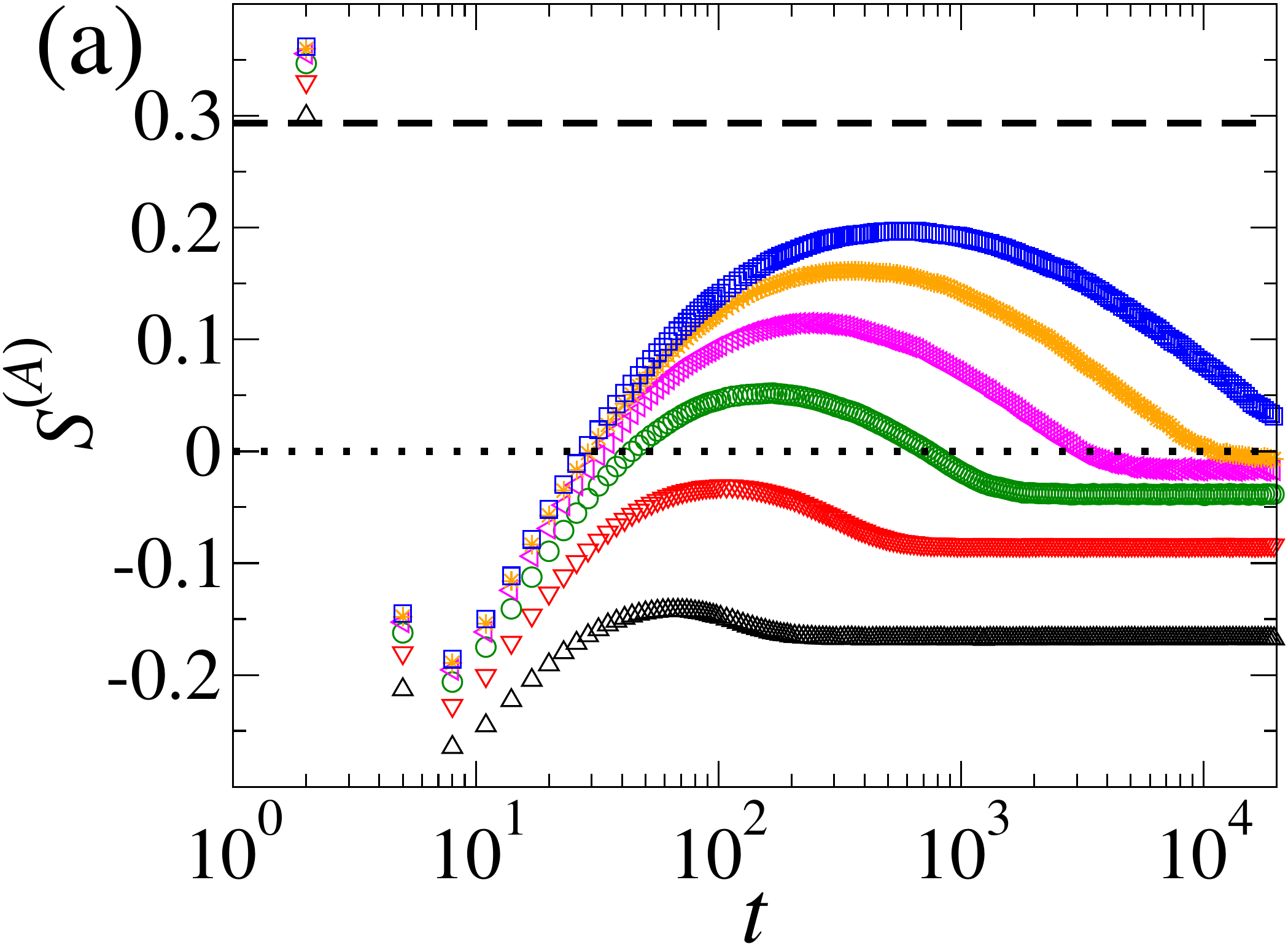}
	\includegraphics[width=4.25cm]{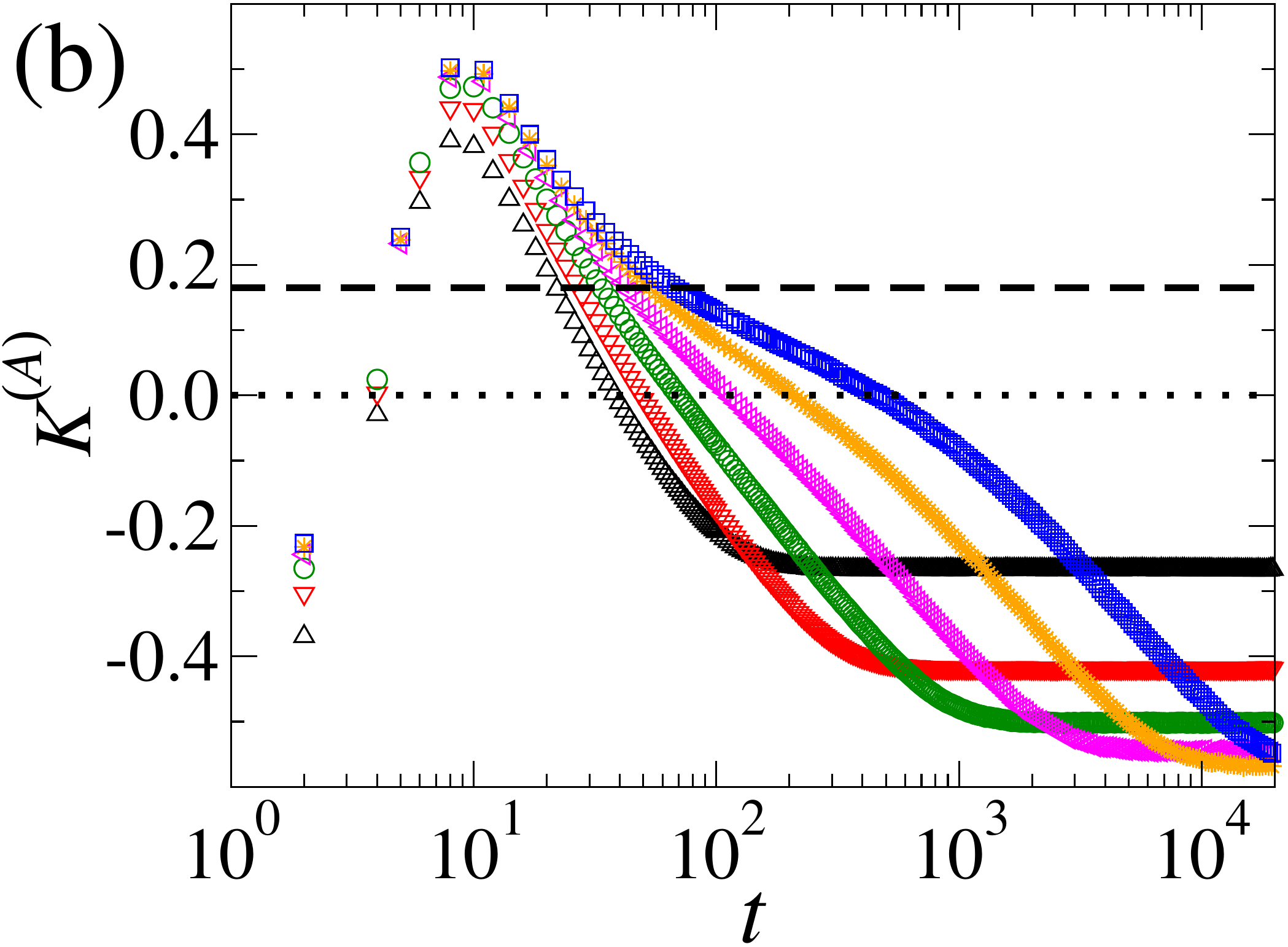}
	\includegraphics[width=4.25cm]{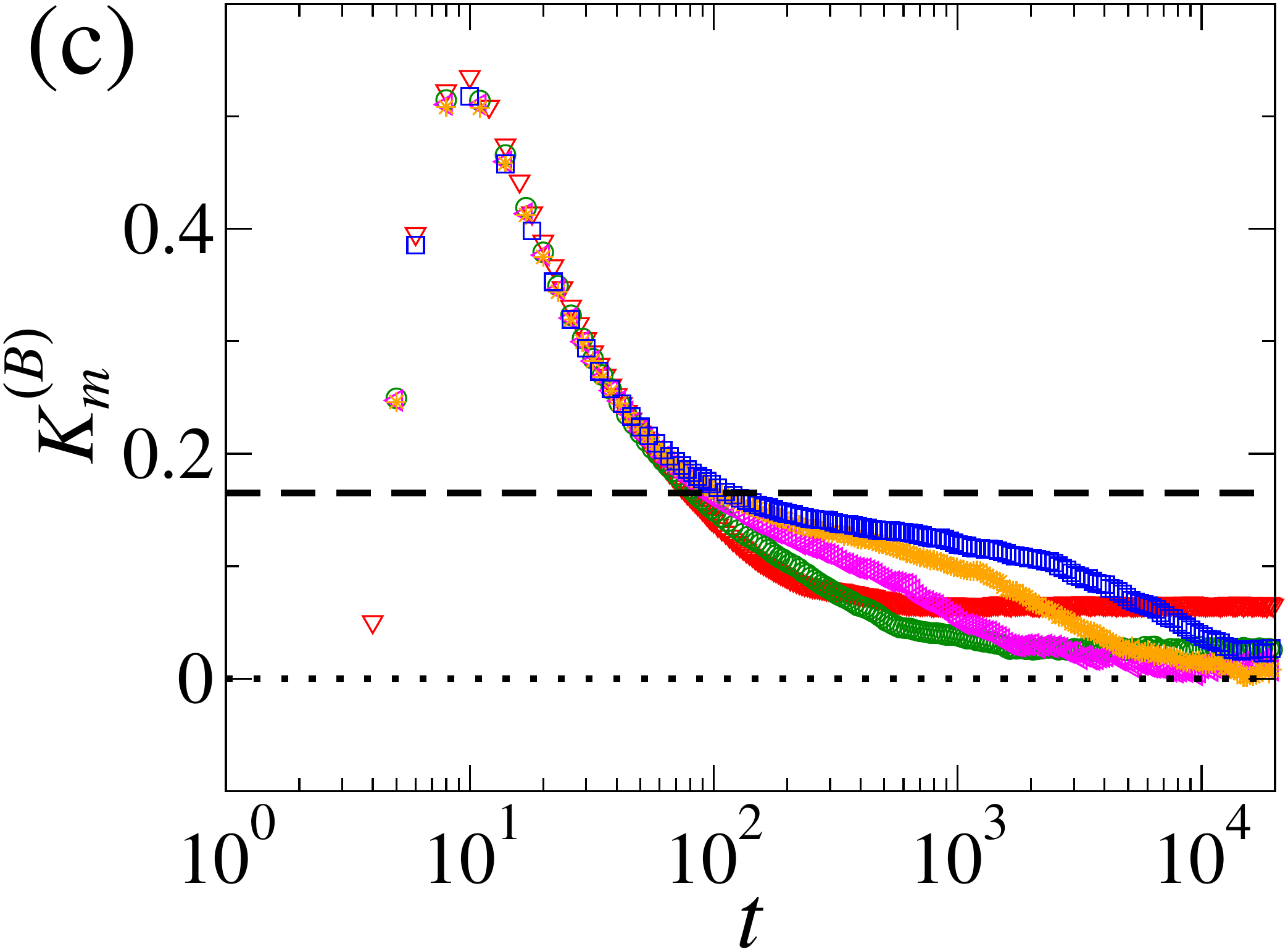}
	\includegraphics[width=4.25cm]{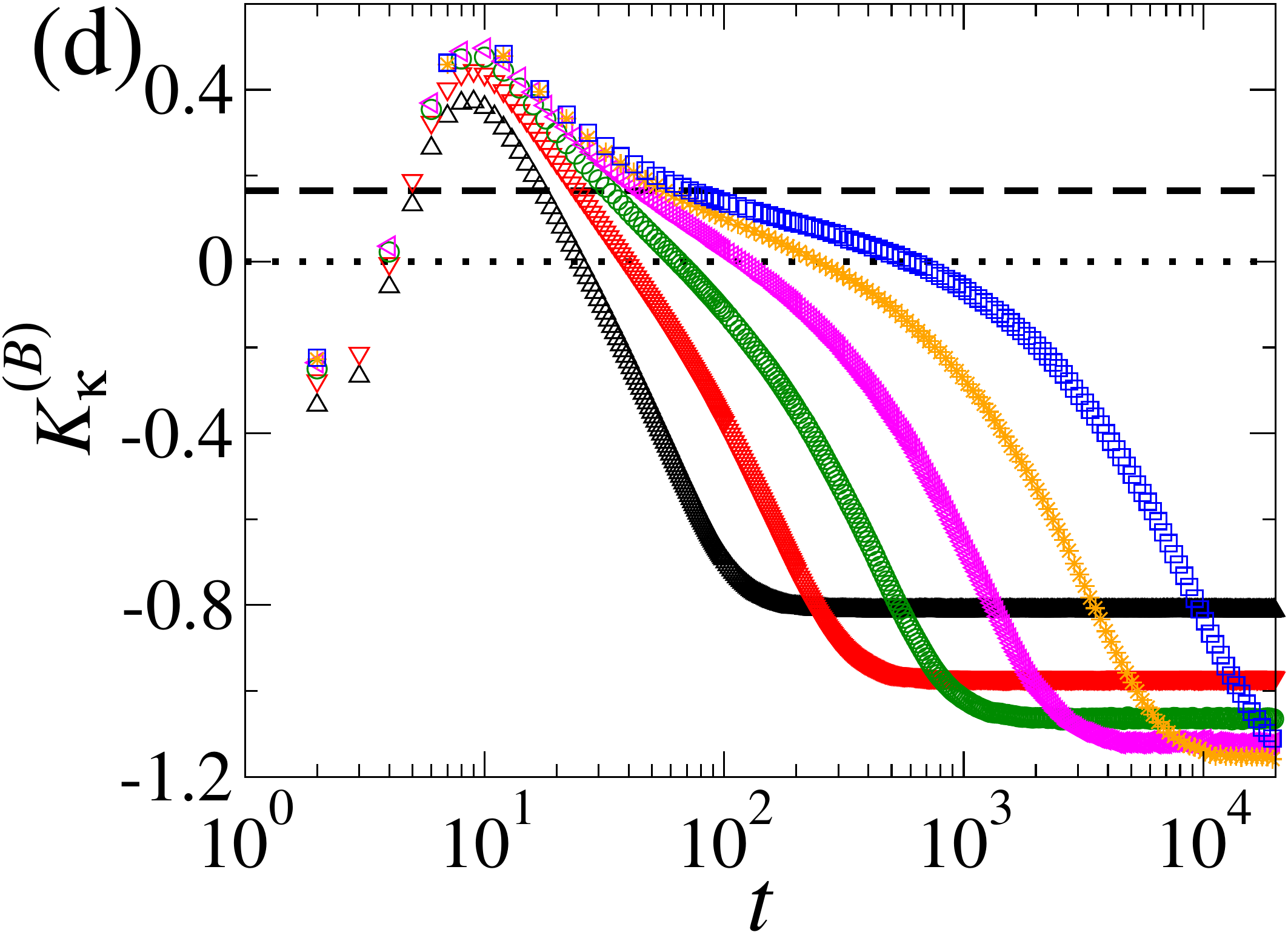}
	\includegraphics[width=4.25cm]{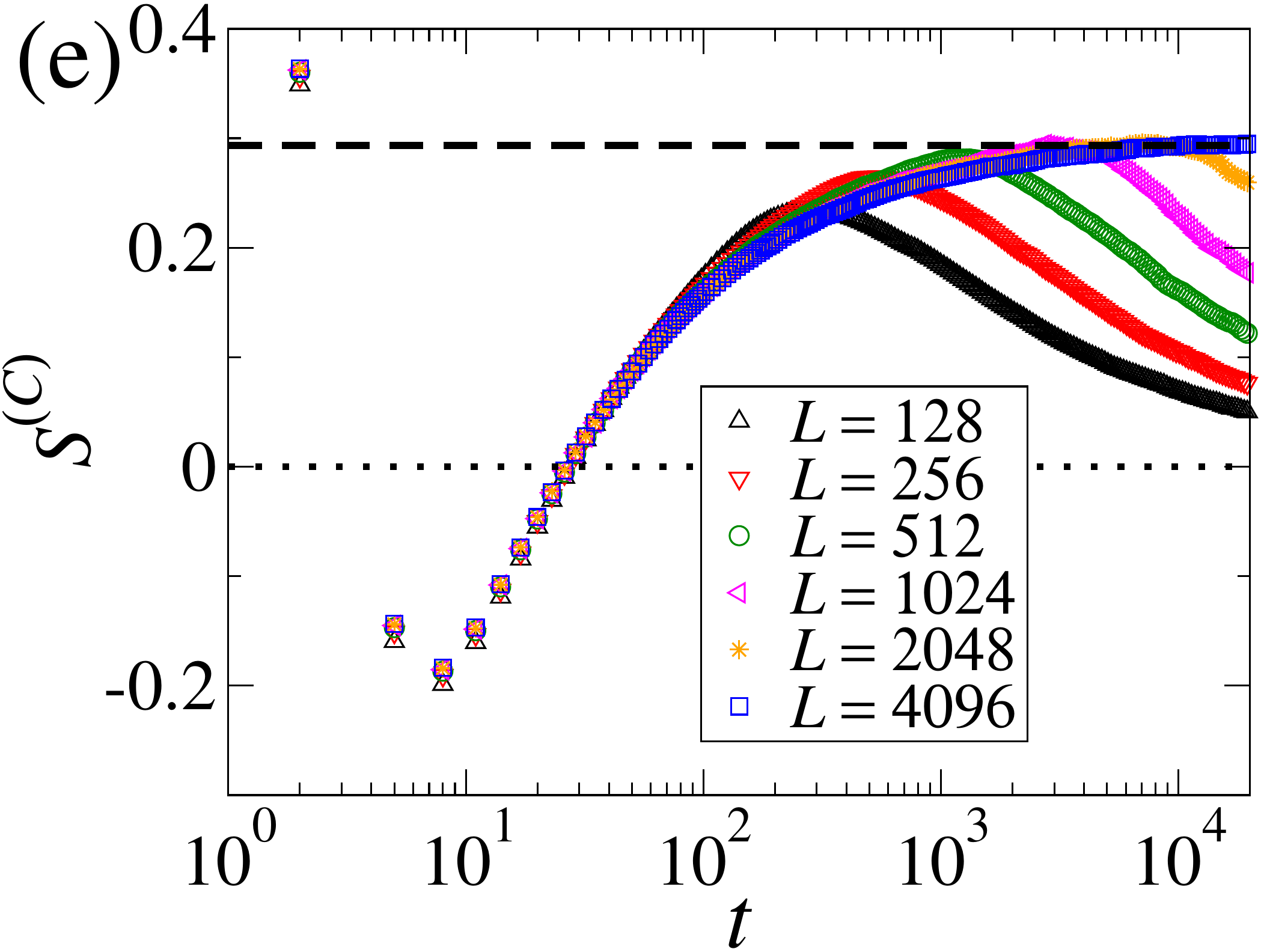}
	\includegraphics[width=4.25cm]{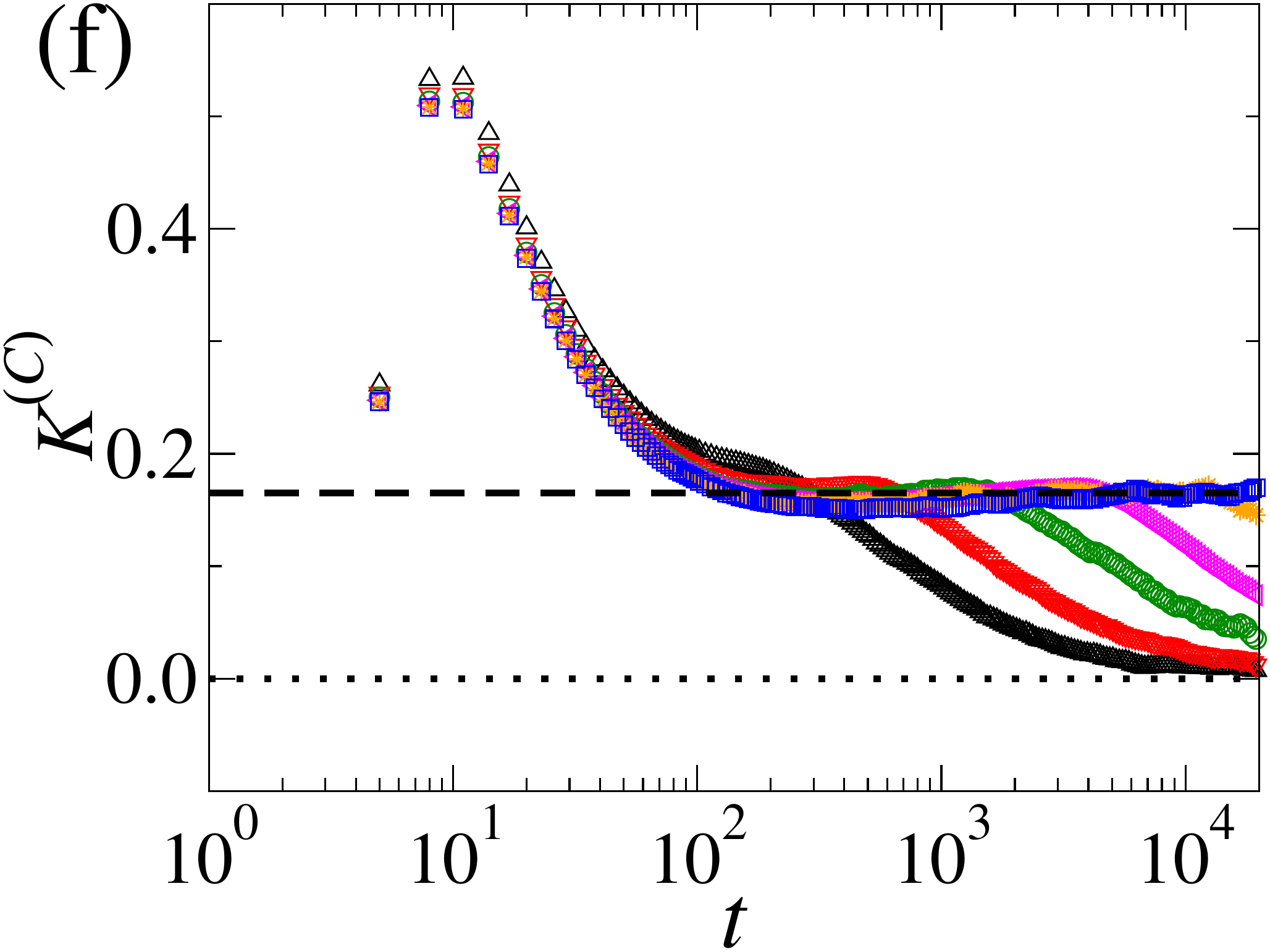}
	\caption{(a)-(f) Temporal variation of the skewness $S^{(x)}$ and kurtosis $K^{(x)}$, with $x=A,B,C$, for the BD model on 1D substrates of several sizes, as indicated in panel (e). The dashed (dotted) horizontal lines represent the ratios for the TW-GOE (Gaussian) distribution.}
	\label{fig3}
\end{figure}

The temporal evolutions of $S$ and $K$ for the 1D BD model are displayed in Fig. \ref{fig3}. The behavior of $S$ is qualitatively the same found for the RSOS model in Fig. \ref{fig2}, with maxima approximating $S_{GOE}$ as $L$ increases in all cases. [Results for $S^{(B)}$ are similar to those in Fig. \ref{fig3}(a) for $S^{(A)}$.] It turns out that in BD case the finite-size effects are more severe than in the RSOS model, as it is widely known (see, e.g., Ref. \cite{Alves14BD} for a survey of literature). For instance, for the smaller substrates analyzed here, the maxima in the $S^{(A)}$ [and also in $S^{(B)}$] curves have negative values, although the BD surfaces are expected to have positive skewness, as indeed observed for larger $L$'s. Moreover, these maxima are still far from $S_{GOE}$ even for  the largest $L$'s considered here. For approach $(C)$, however, one always finds positive maxima and for $L \geqslant 512$ they are quite close to the expected TW-GOE value. So, once again, this last approach returns much better results than the other ones. This is even more evident in the kurtosis, for which no signature of TW-GOE statistics is obtained from methods $(A)$ and $(B)$, for the sizes analyzed here, beyond a shoulder in the $K$ \textit{vs.} $t$ curves depicted in Fig. \ref{fig3}. In case $(C)$, notwithstanding, one finds approximated plateaus around $K_{GOE}$ already for $L \geqslant 256$. It is quite remarkable that for a difficult model as the ballistic deposition the ``1-pt statistics'' is capable of uncovering the asymptotic HD properties for so small substrate sizes and times.

\begin{figure}[!t]
	\includegraphics[width=4.25cm]{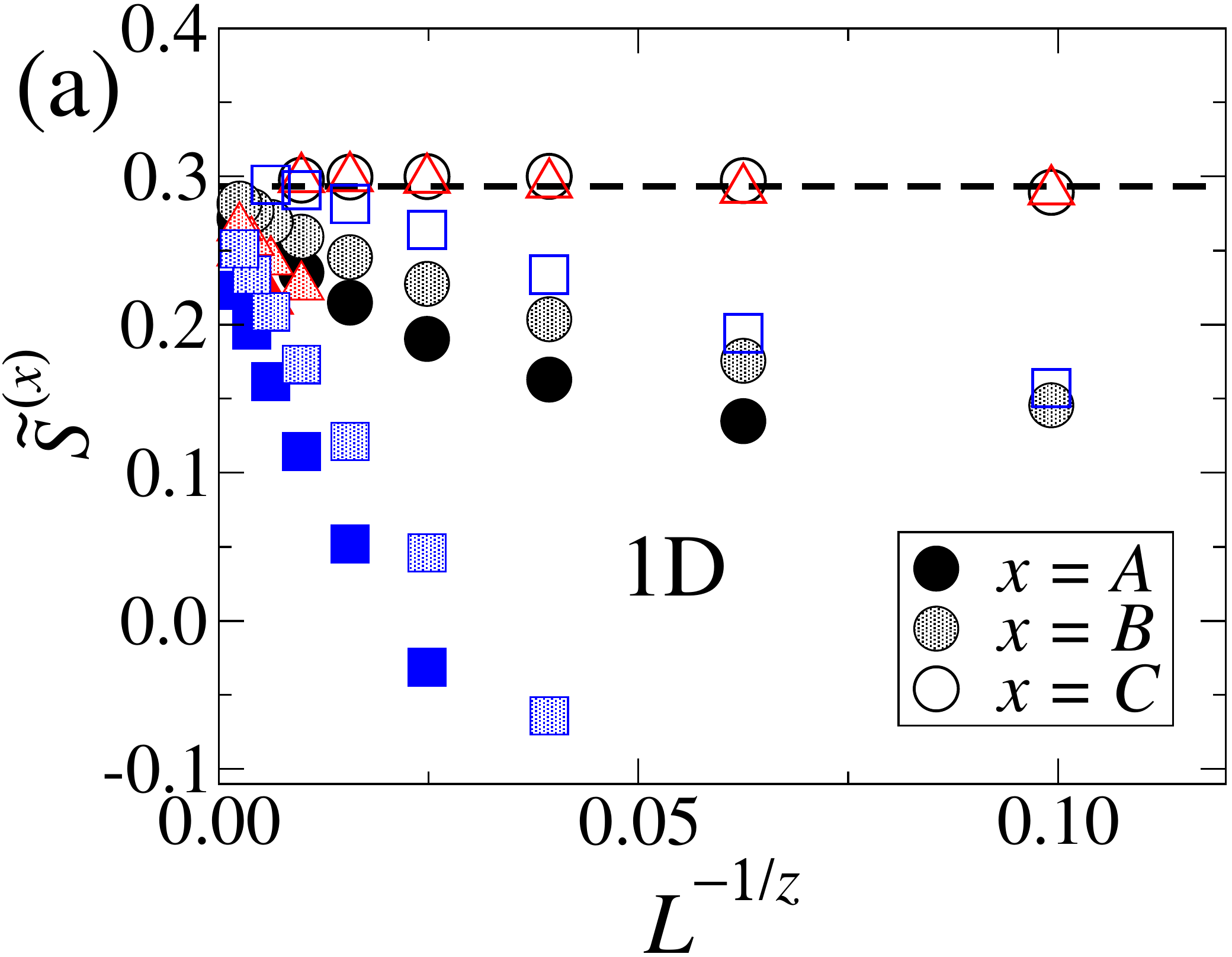}
	\includegraphics[width=4.25cm]{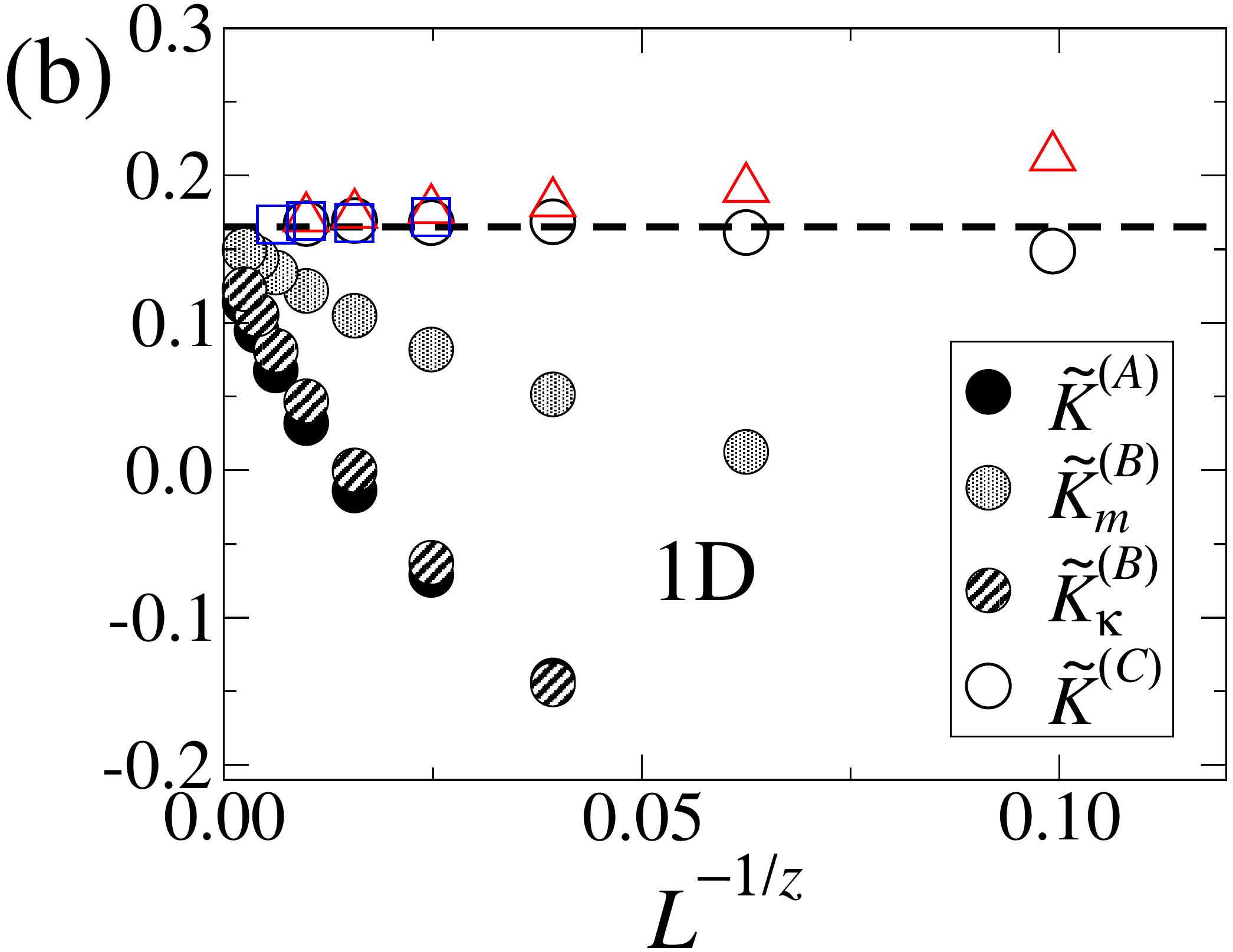}
	\includegraphics[width=4.25cm]{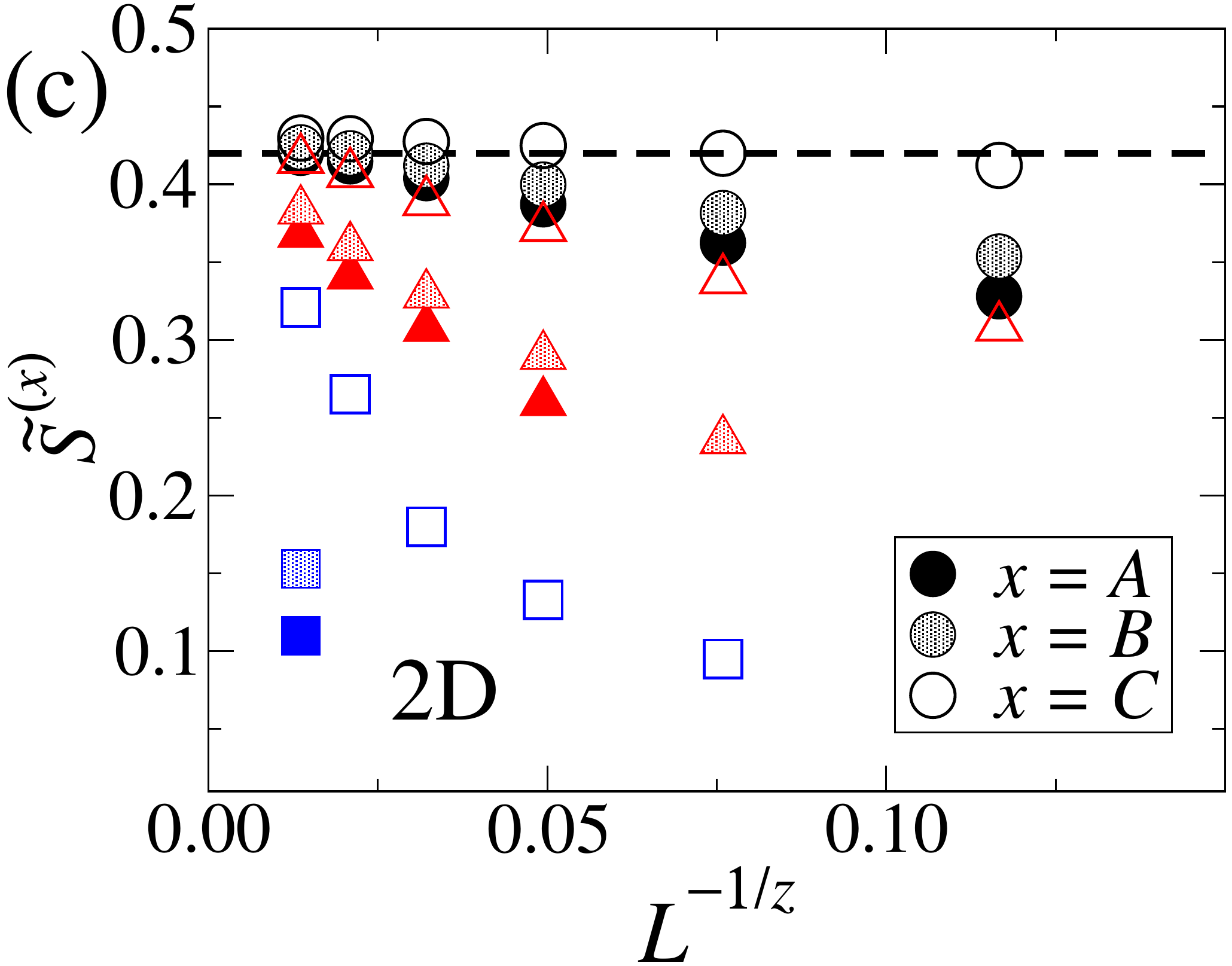}
	\includegraphics[width=4.25cm]{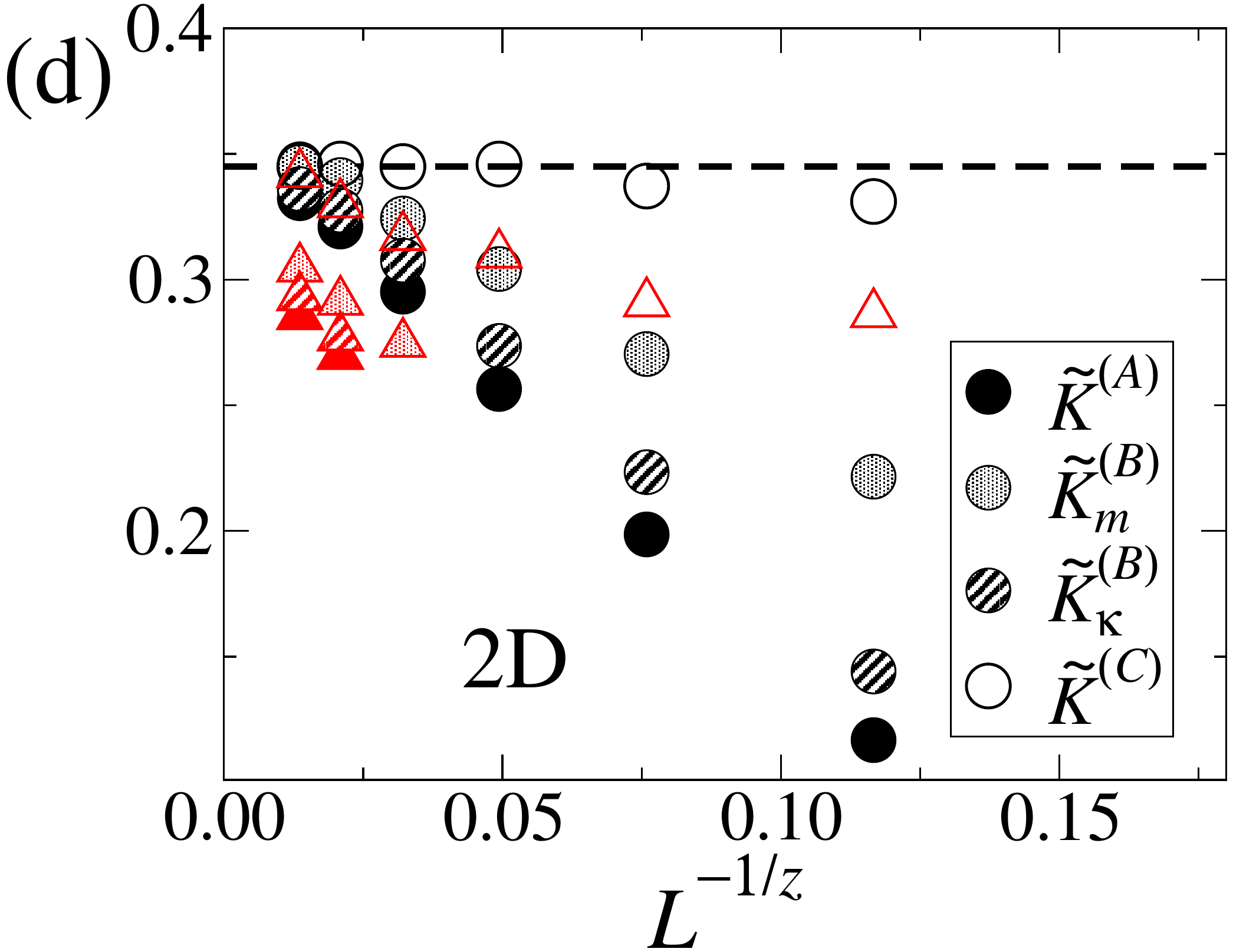}
	\caption{Maxima in the skewness $\tilde{S}$ (left) and kurtosis $\tilde{K}$ (right panels) versus $L^{-1/z}$, for the RSOS (black circles), BD (blue squares) and Etching model (red triangles). Data for different approaches are shown, as indicated by the legends. The top panels show results for 1D substrates, where $z=3/2$ and the dashed horizontal lines represent the values of $S_{GOE}$ and $K_{GOE}$. Results for 2D substrates are displayed in the bottom panels, where $z=1.613$ \cite{Pagnani} is used and the dashed lines are the best numerical estimates for the asymptotic ratios ($S \simeq 0.425$ and $K \simeq 0.346$ \cite{healy12,tiago13}) for the 2D KPZ HDs in the GR.}
	\label{fig4}
\end{figure}

Results (not shown) for the 1D Etching model are somewhat intermediate between those for the RSOS and BD model. For example, in the skewness estimated from approaches $(A)$ and $(B)$, the maxima associated with the asymptotic GR HDs appear only for $L\geqslant 2048$. Namely, the situation worsens when compared with the BD case. However, the corrections in $S^{(C)}$ are smaller than in the BD case, being comparable to those in the RSOS model. In the kurtosis, everything is similar to the BD model, but once again with milder finite-size effects in $K^{(C)}$. This is demonstrated in Figs. \ref{fig4}(a) and \ref{fig4}(b), where estimates of $\tilde{S}$ and $\tilde{K}$ are depicted against $L^{-1/z}$, with $z=3/2$ being the dynamic exponent of the 1D KPZ class \cite{KPZ}. These maxima let clear that, for a given $L$, corrections in $S^{(A)}$ are more severe than in $S^{(B)}$, while in $S^{(C)}$ they are quite small, as expected from the results above. In the kurtosis [Fig. \ref{fig4}(b)], one observes similar corrections in $K^{(A)}$ and $K_{\kappa}^{(B)}$, which are stronger than those in $K_m^{(B)}$ for the RSOS model. Note that for the other models no maximum related to the asymptotic HD was found in the curves of $K \times  t$ for approaches $(A)$ and $(B)$, for the sizes analyzed here. 

Despite the different corrections, in all cases the maxima converge toward $S_{GOE}$ and $K_{GOE}$ as $L$ increases. According to Ref.  \cite{LandauShim}, the dominant correction in both $\tilde{S}^{(A)}$ and $\tilde{K}^{(A)}$ is of type $L^{-1/z}$, so that $\tilde{F}(L)=F_{\infty} + a_F L^{-1/z}$, with $F=S$ or $K$. This is the reason for the abscissa $L^{-1/z}$ in Fig. \ref{fig4}, but for the sizes considered here ($L \leqslant 8192$) I do not find clear linear behaviors there. [Data for $L$ up to $10^6$ was considered in Ref. \cite{LandauShim}, which certainly explains the difference.] I notice that by considering also multiplicative logarithmic corrections, so that $\tilde{F}(L)=F_{\infty} + a_F L^{-1/z} [\log(1/L)]^{\gamma}$, with $\gamma=5/3$ for $F=S$ and $\gamma=3/2$ for $F=K$, the data for the RSOS model are well-linearized, even for small $L$'s, and this returns extrapolated values quite close to the expected $S_{GOE}$ and $K_{GOE}$ ones.

Now, I will discuss results for the KPZ models deposited on flat 2D substrates.

In general, the temporal variations of $S$ and $K$ in 2D are very similar to those found for the 1D case [in Figs. \ref{fig1}, \ref{fig2} and \ref{fig3}], but with stronger finite-size corrections. In the BD model such corrections are actually very severe, so that the maxima observed in $S^{(A)}$ and $S^{(B)}$ in the 1D case [see Fig. \ref{fig3}(a)] only appear for $L\gtrsim 1024$ in $d=2$. In the same fashion, only for such large sizes shoulders start appearing in the curves of $K$ \textit{vs.} $t$, for procedures $(A)$ and $(B)$. Therefore, for these two approaches, no signature of the asymptotic GR HD is found in the ratios for the 2D BD model with $L \lesssim 1024$. However, maxima are always found in $S^{(C)}$ and the shoulder appears for smaller $L$'s in $K^{(C)}$ curves for the BD model. For the sake of conciseness, I will omit such curves here, passing directly to the analysis of their maxima, whenever they exist, which are depicted in Figs. \ref{fig4}(c) and \ref{fig4}(d). Similarly to 1D case, one sees in these figures that the deviations from the expected asymptotic ratios, for a given $L$, decreases as one goes from approach $(A)$ to $(B)$ and from $(B)$ to $(C)$. This confirms that method $(A)$ is the worse to analyze the GR HDs, while procedure $(C)$ provides the best results. This is particularly notable for the RSOS model, where $\tilde{S}^{(C)}$ and $\tilde{K}^{(C)}$ agree quite well with the asymptotic values already for $L \gtrsim 128$. For comparison, substrates with $32768\times32768$ sites and temporal extrapolations were used to access these asymptotic values in Refs. \cite{tiago13,Ismael14}, via the method ($B$) for the cumulants.

\subsection{KPZ HDs for the steady state regime}
\label{subsecSSR}

Next, I investigate the height fluctuations for the regime where the interfaces are fully correlated. While in the GR HDs analyzed in the previous subsection only the finite-size and -time corrections are affected by the method used to estimate $S$ and $K$, in the SSR different asymptotic values (for $L \rightarrow \infty$) are obtained, depending on the way that these ratios are calculated. This is clear in Figs. \ref{fig1}-\ref{fig3}, where one sees that $S$ and $K$ attain constant values at long times for methods $(A)$ and $(B)$, but with $K^{(A)} \neq K_m^{(B)} \neq K_{\kappa}^{(B)}$. However, $S^{(C)}$ and $K^{(C)}$ do not saturate; instead, they present a slow decrease in magnitude for the same times where the ratios from approaches $(A)$ and $(B)$ are already constant (for a given $L$). The very same thing happens in the 2D case.

\begin{figure}[!t]
	\includegraphics[width=4.25cm]{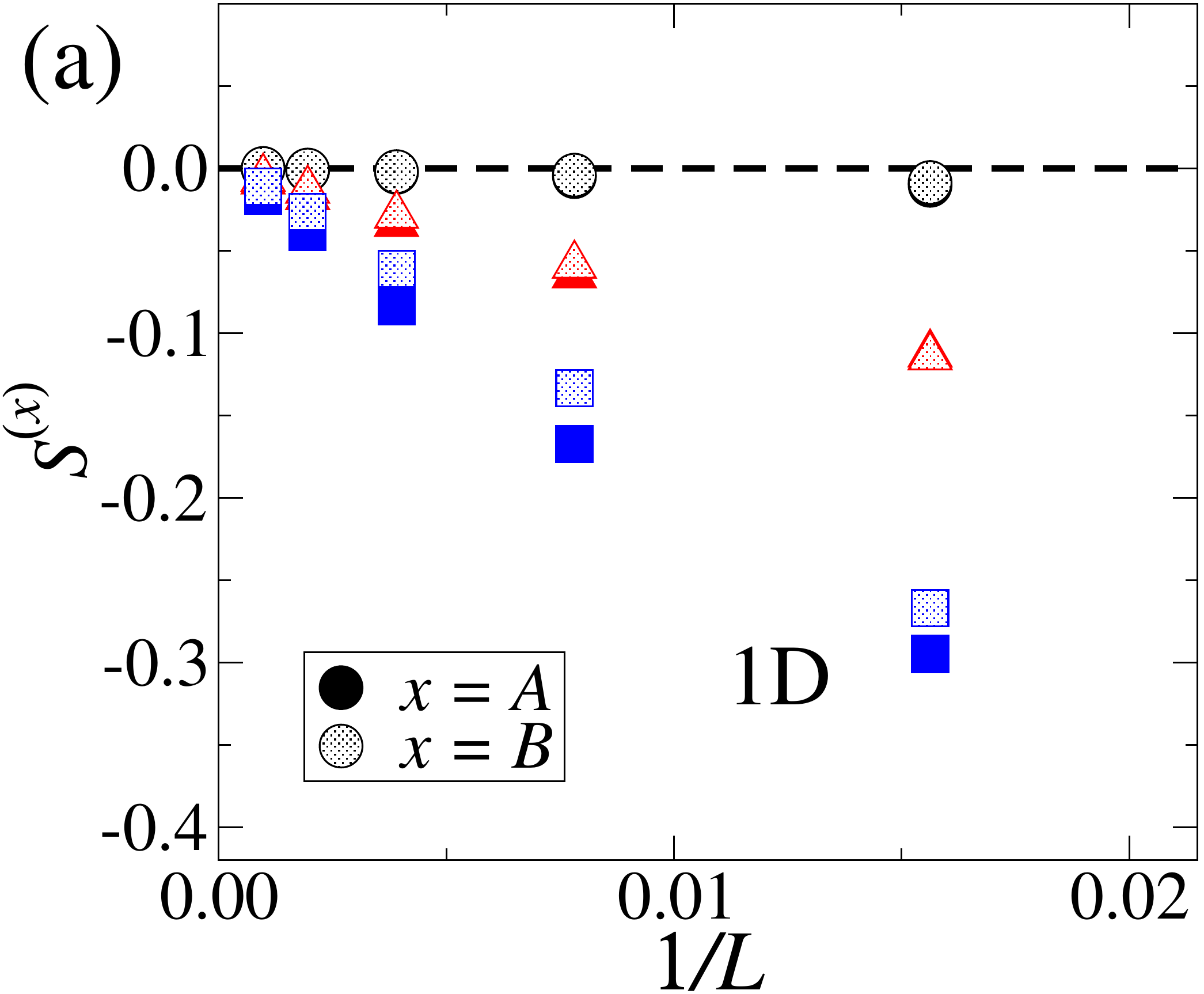}
	\includegraphics[width=4.25cm]{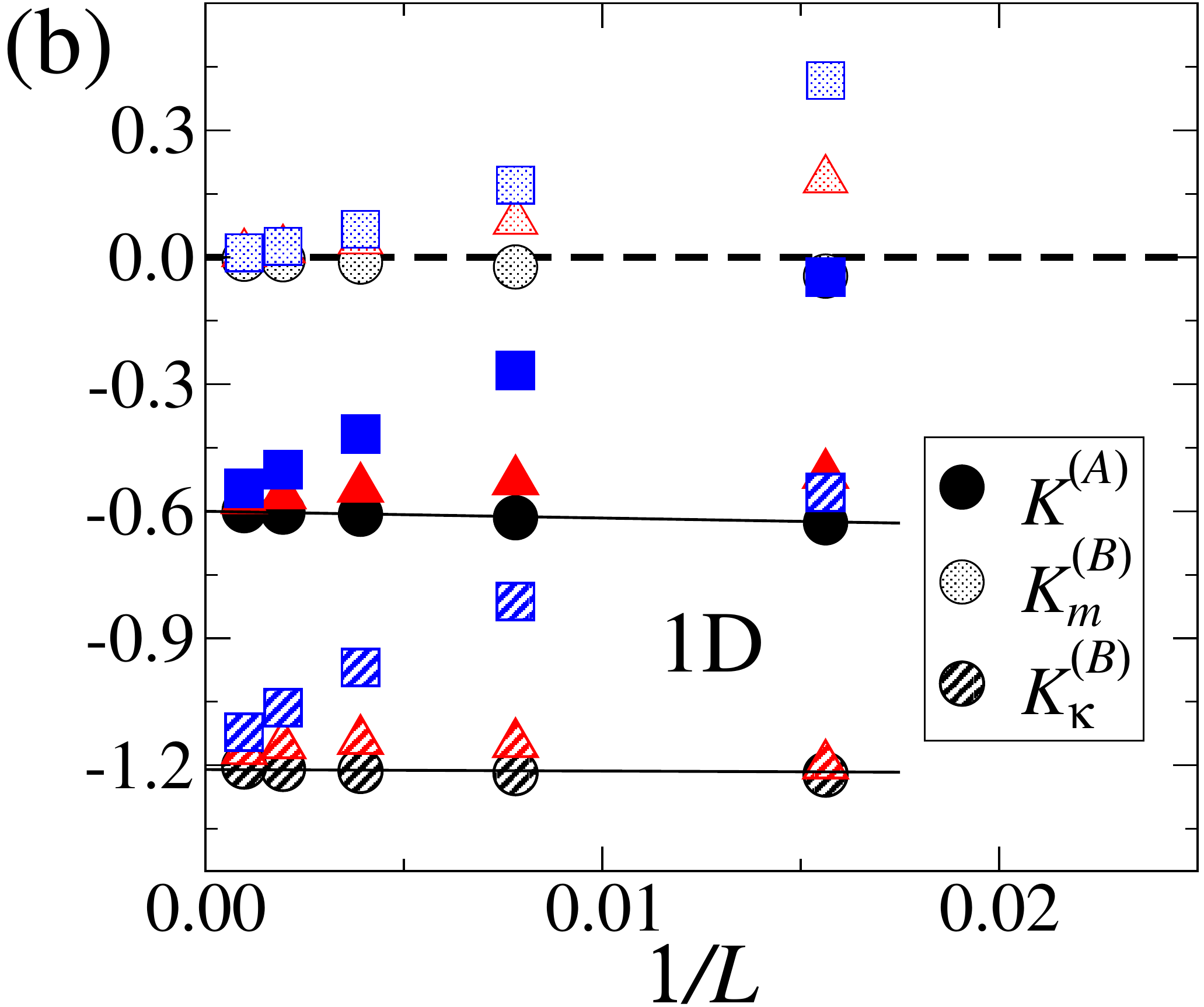}
	\includegraphics[width=4.25cm]{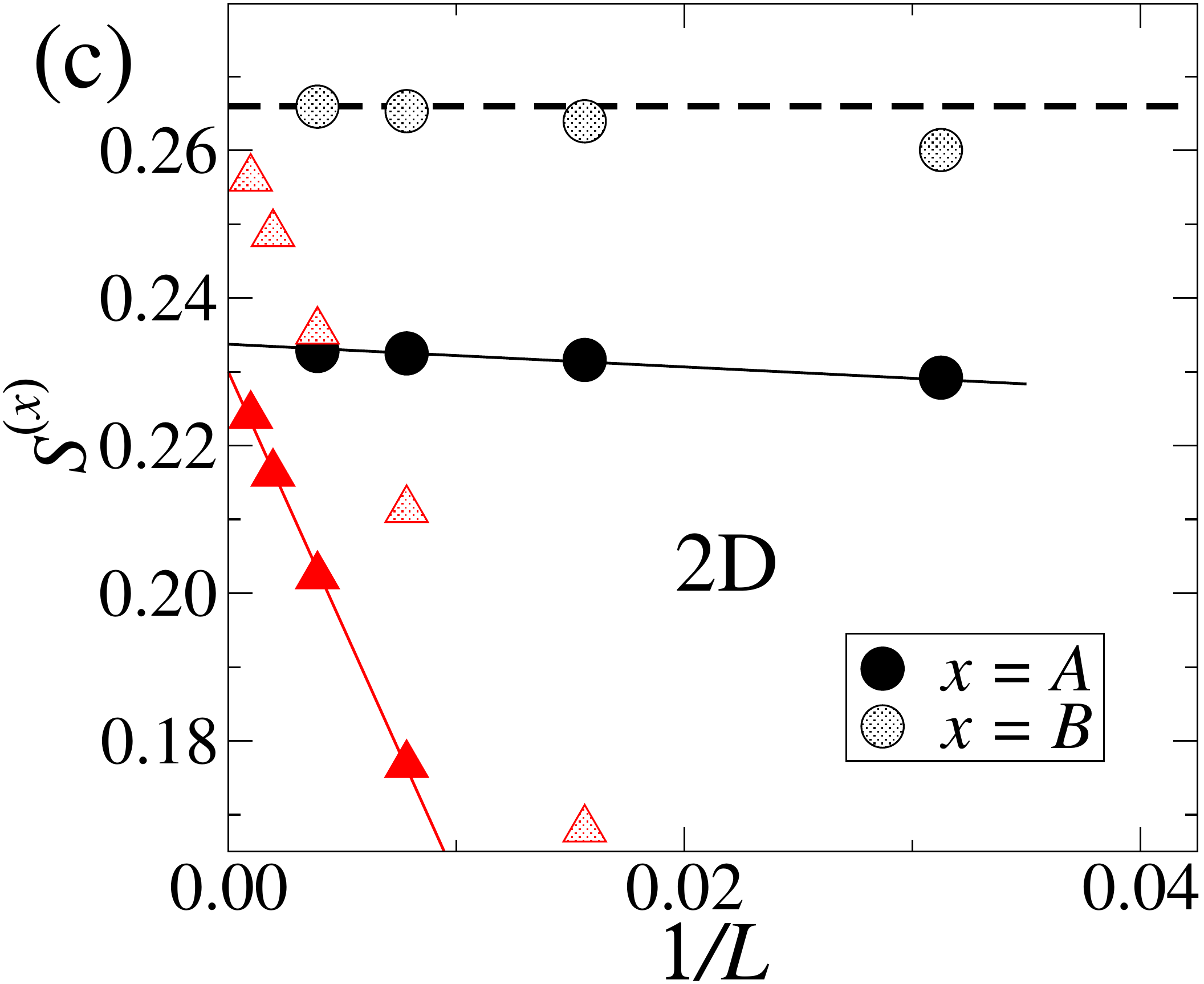}
	\includegraphics[width=4.25cm]{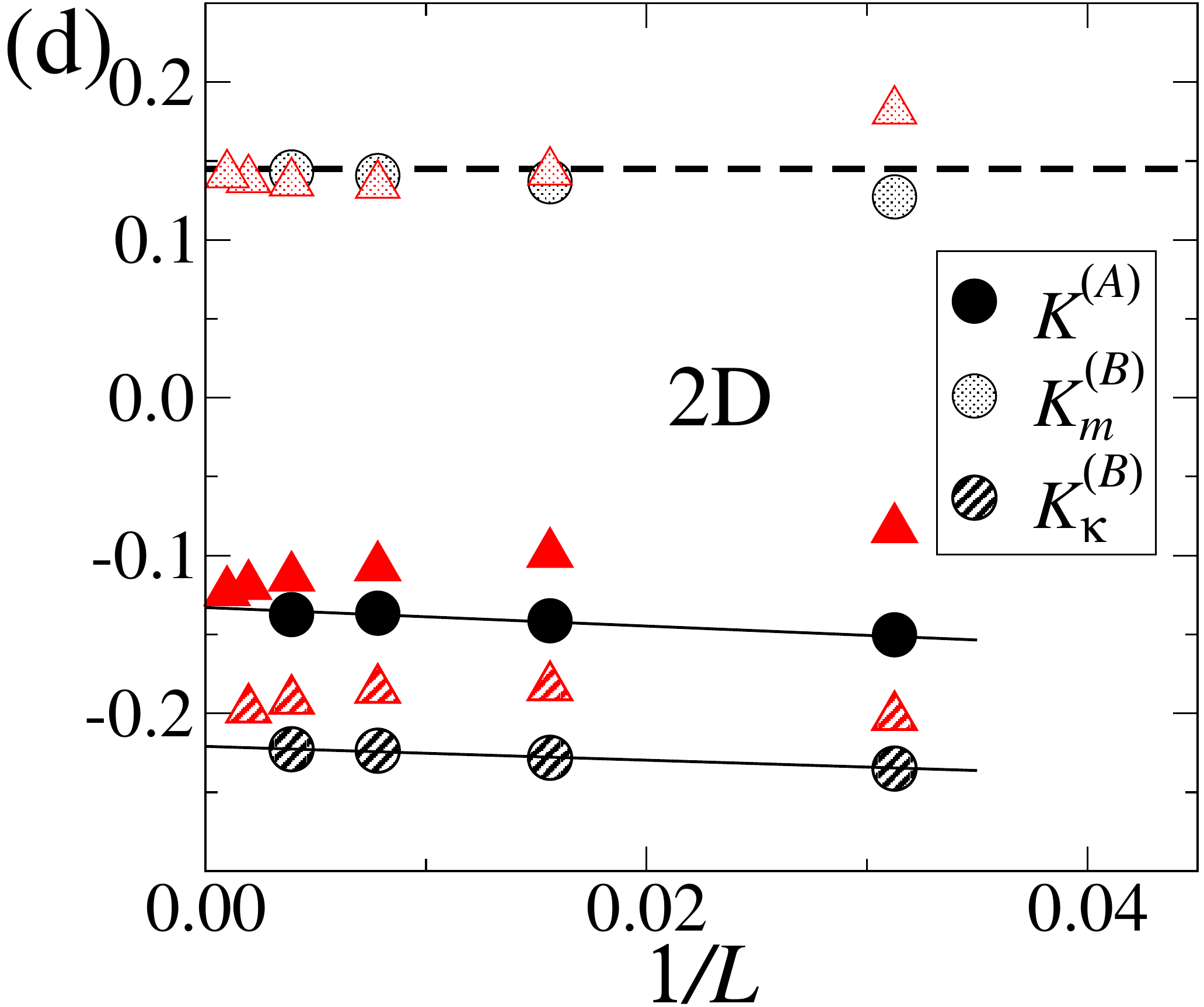}
	\caption{Saturation values of the skewness $S$ (left) and kurtosis $K$ (right panels) versus $1/L$, for the RSOS (black circles), BD (blue squares) and Etching model (red triangles). Data for different approaches are shown, as indicated by the legends. The top (bottom) panels show results for 1D (2D) substrates, where the dashed horizontal lines represent the values of the central moment ratios: $S=K=0$ in 1D and $|S|\simeq 0.266$ and $K \simeq 0.145$ \cite{Pagnani} in 2D case. The solid lines are linear fits used in extrapolations. For the RSOS model, $-S$ is shown. Data for the BD model are omitted in panels (c) and (d) because, due to the strong finite-size corrections, most points are out of the vertical range shown.}
	\label{fig5}
\end{figure}

Estimates for the saturation values of $S^{(A)}$ and $S^{(B)}$ are depicted in Fig. \ref{fig5}(a) against $1/L$ for the three 1D KPZ models. Although the results for method $(A)$ display slightly stronger finite-size corrections in some cases, in both approaches and for all models the skewness vanishes as $L$ increases, as expected for interfaces with up-down symmetry \cite{barabasi}. Figure \ref{fig5}(c) presents a similar plot for the KPZ models on 2D substrates, where two (close, but) different asymptotic values are found for the skewness. In fact, while $|S^{(B)}|$ converges to close to $0.266$, in agreement with the finding by several authors \cite{Chin,Marinari,Fabio2004kpz,Pagnani}, in approach $(A)$ the skewness extrapolates to $|S^{(A)}| = 0.232(3)$. This last value agrees with $|S| = 0.23(2)$ reported in Ref. \cite{LandauShim}, while they both differ from $|S^{(B)}|$. Hence, different values of $S$ can be obtained for the same ensemble of interfaces depending on the way this ratio is calculated.

The situation ``worsens'' in the kurtosis, for which a diversity of asymptotic values can be found. For the 1D case, one sees in Fig. \ref{fig5}(b) that the saturation value of $K_m^{(B)}$ vanishes, as expected for a Gaussian distribution. However, the $K$'s for the other approaches clearly converge to other values. The extrapolated value of $K^{(A)}$ for the RSOS model is $K^{(A)} = -0.592(4)$, which agrees with less accurate extrapolations for the other models, as well as with the value reported in Ref. \cite{LandauShim} [$K = -0.61(2)$] within the error bars. The data for $K_{\kappa}^{(B)}$ are harder to extrapolate, for the sizes analyzed here. For instance, they seem to present logarithmic corrections in the RSOS model and display a non-monotonic convergence in the Etching case. Notwithstanding, one may observe in Fig. \ref{fig5}(b) that they are converging toward $K_{\kappa}^{(B)} \approx -1.20$ as $L$ increases. Different asymptotic values are obtained also in the 2D case, with $K^{(A)} \approx -0.134$, $K_{\kappa}^{(B)} \approx -0.22$ and $K_{m}^{(B)} \approx 0.14$ [see Fig. \ref{fig5}(d)]. This last value agrees with those from Refs. \cite{Chin,Marinari,Fabio2004kpz,Pagnani}, while the $K^{(A)}$ found here is consistent with the one estimated in Ref. \cite{LandauShim}. To the best of my knowledge, $K_{\kappa}^{(B)}$ was never reported in the literature.

I recall that the differences between $K_{\kappa}^{(B)}$ and $K_{m}^{(B)}$ are indeed expected, as pointed out in Sec. \ref{secModels}, being a consequence of the width fluctuations. The origins of the differences in the other cases will be carefully discussed in Sec. \ref{secDisc}.

\section{Results for KPZ models on expanding substrates}
\label{secKPZcurved}

As a means of analyzing the differences among the different averaging procedures in curved KPZ interfaces, I investigate the three KPZ models discussed above on 1D substrates enlarging linearly in time, as $\expct{L} = L_0 + \omega t$ \cite{Ismael14}. Since the SSR is never attained in this case, only the GR HD will be discussed here, which for the 1D KPZ class is asymptotically given by the TW-GUE distribution, with $S_{GUE} = 0.22408$ and $K_{GUE} = 0.09345$ \cite{Prahofer2000}.

\begin{figure}[!t]
	\includegraphics[width=4.25cm]{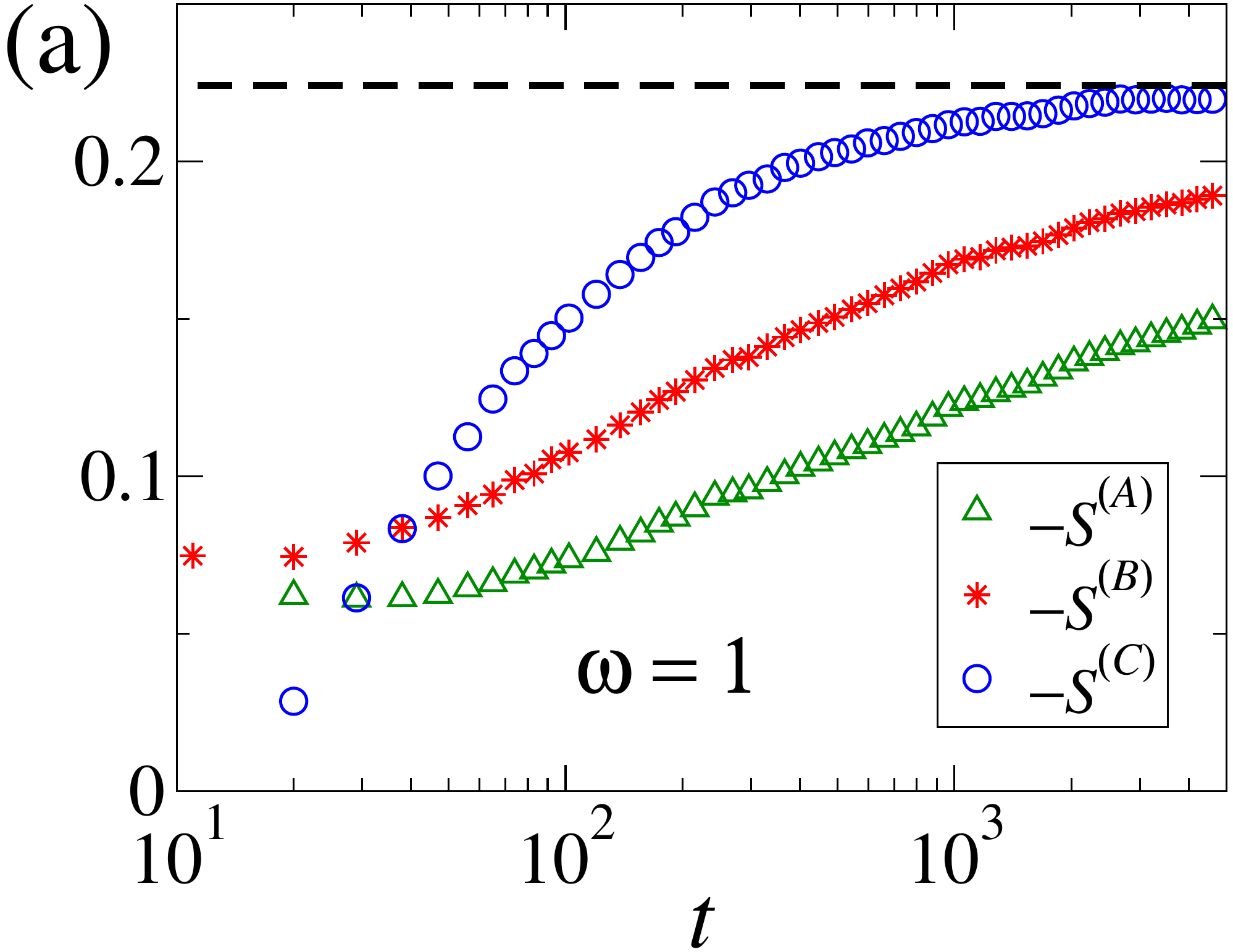}
	\includegraphics[width=4.25cm]{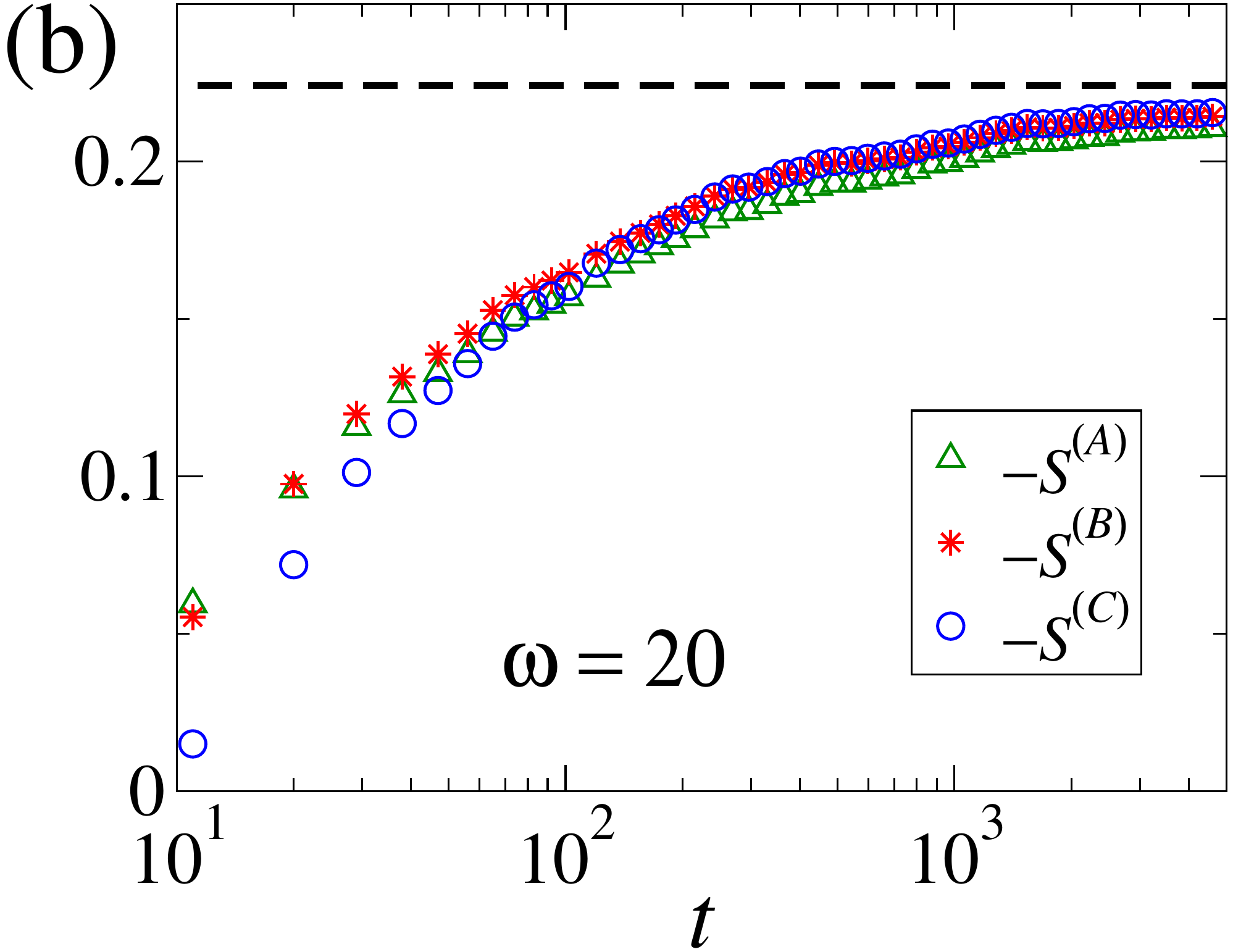}
	\includegraphics[width=4.25cm]{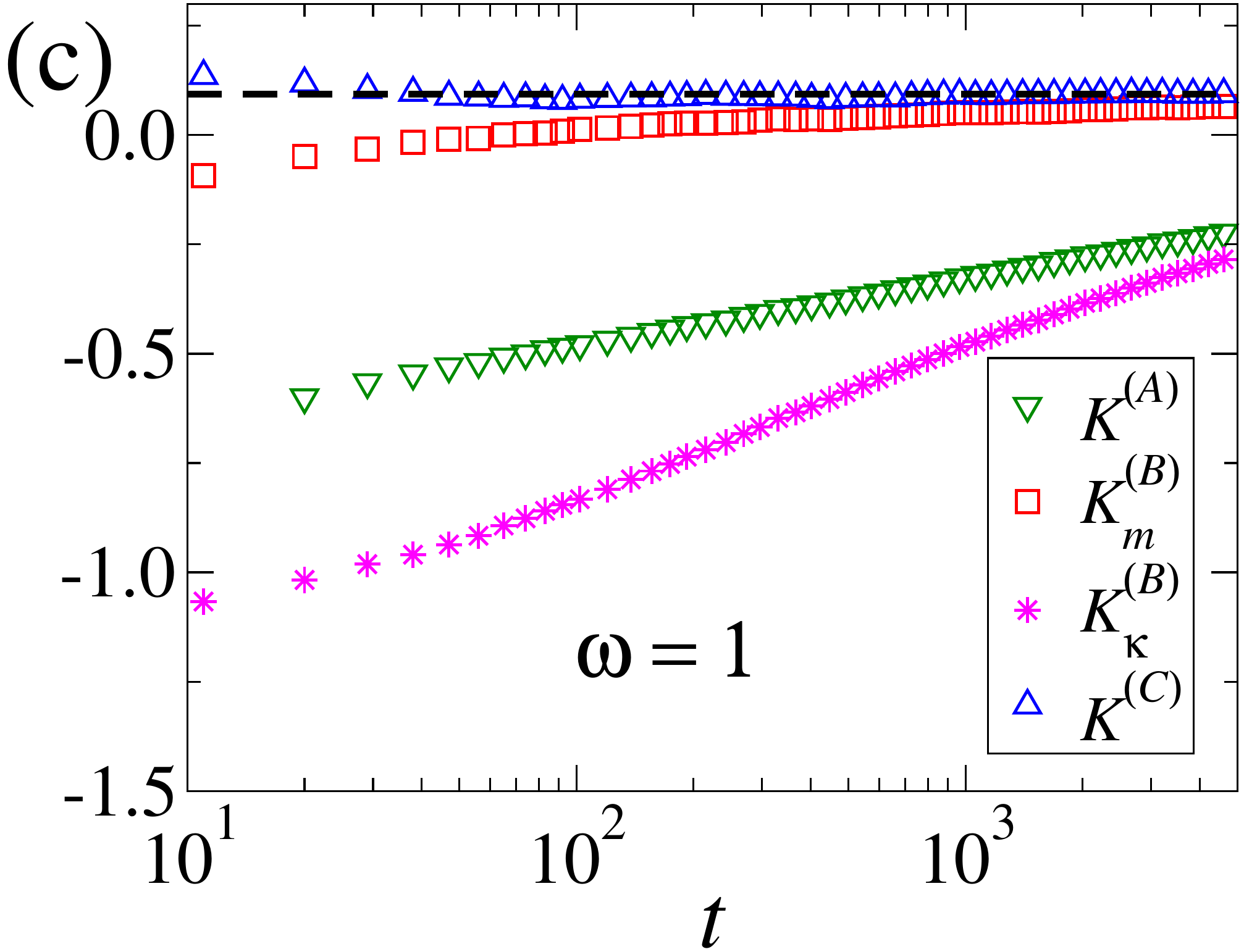}
	\includegraphics[width=4.25cm]{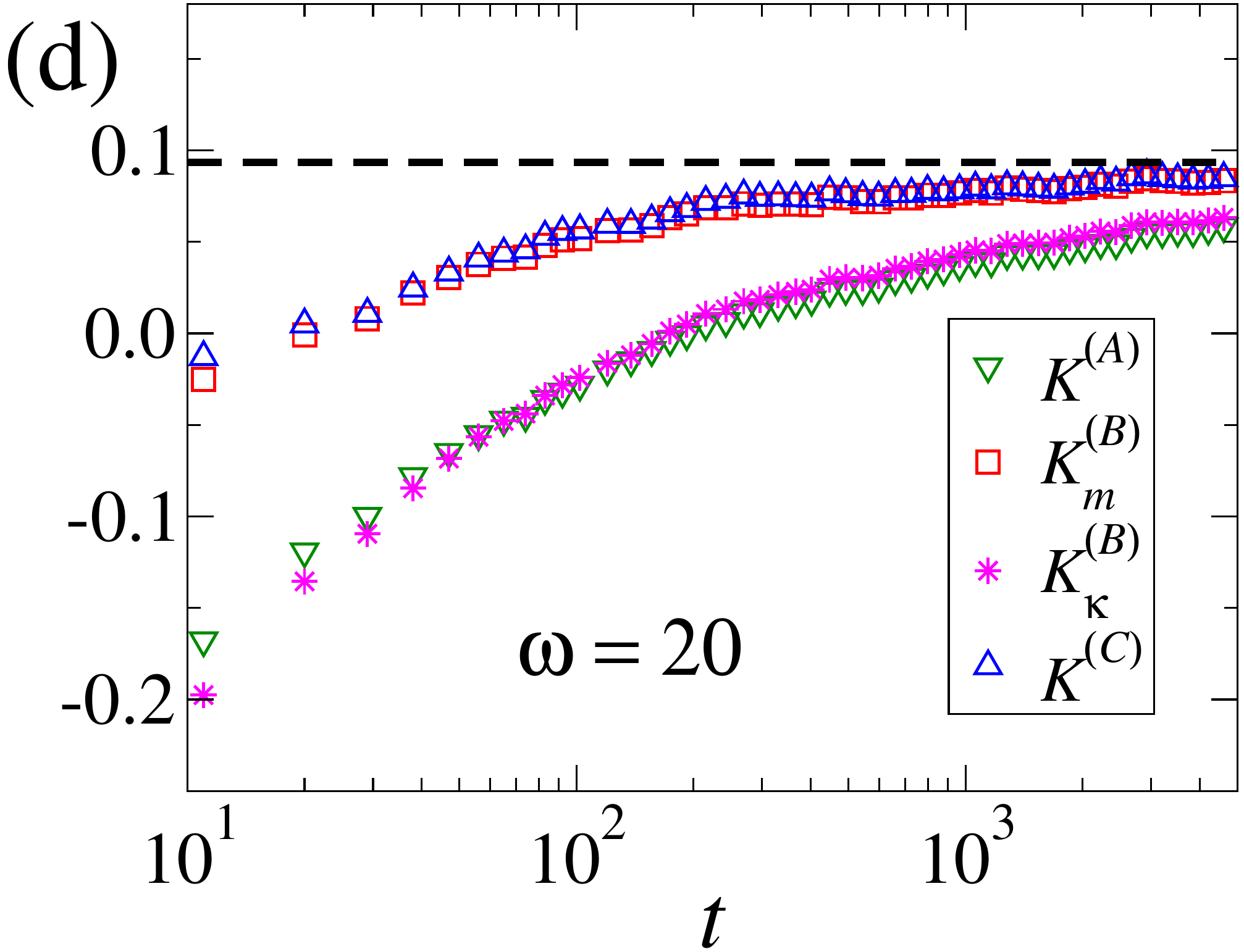}
	\includegraphics[width=4.25cm]{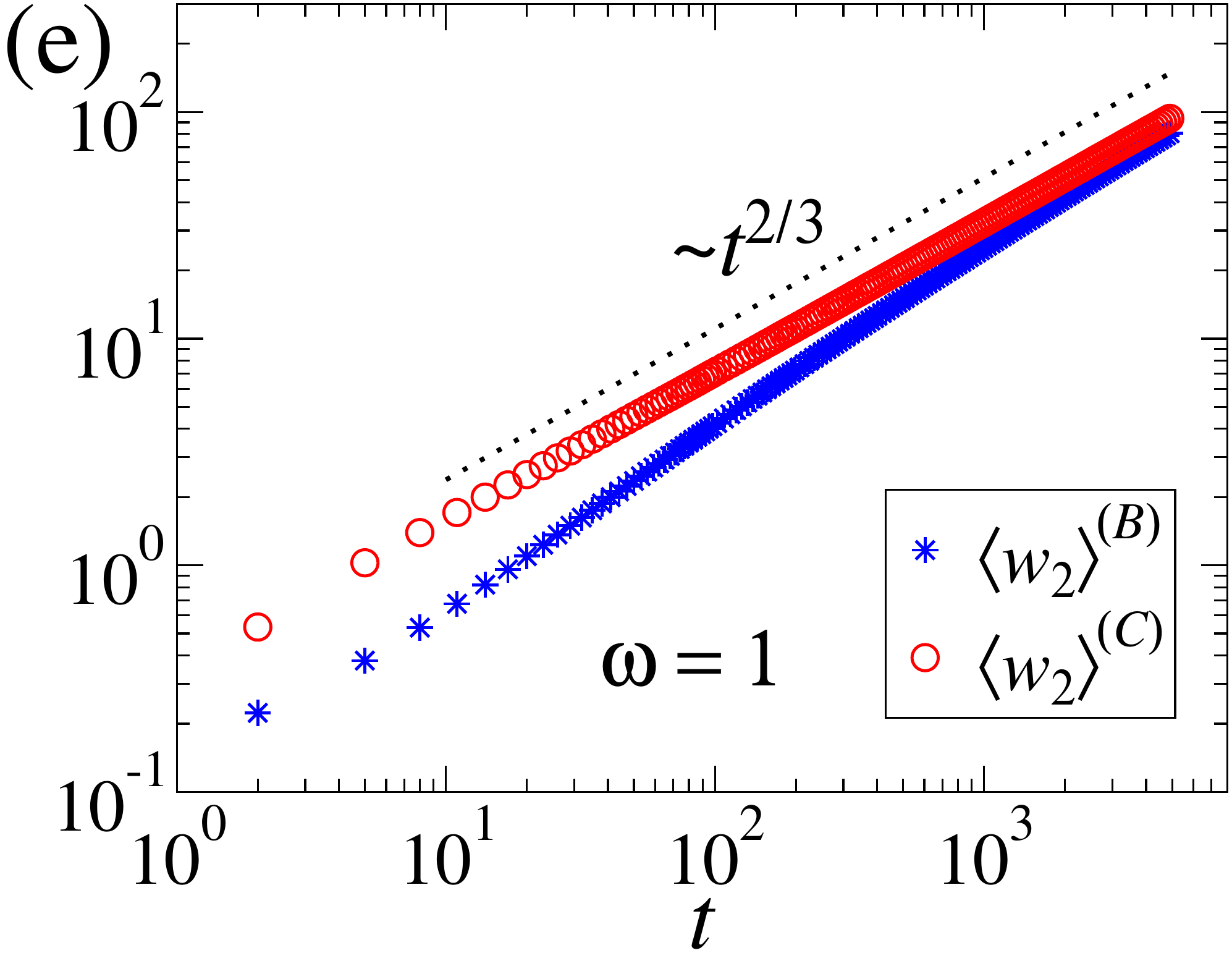}
	\includegraphics[width=4.25cm]{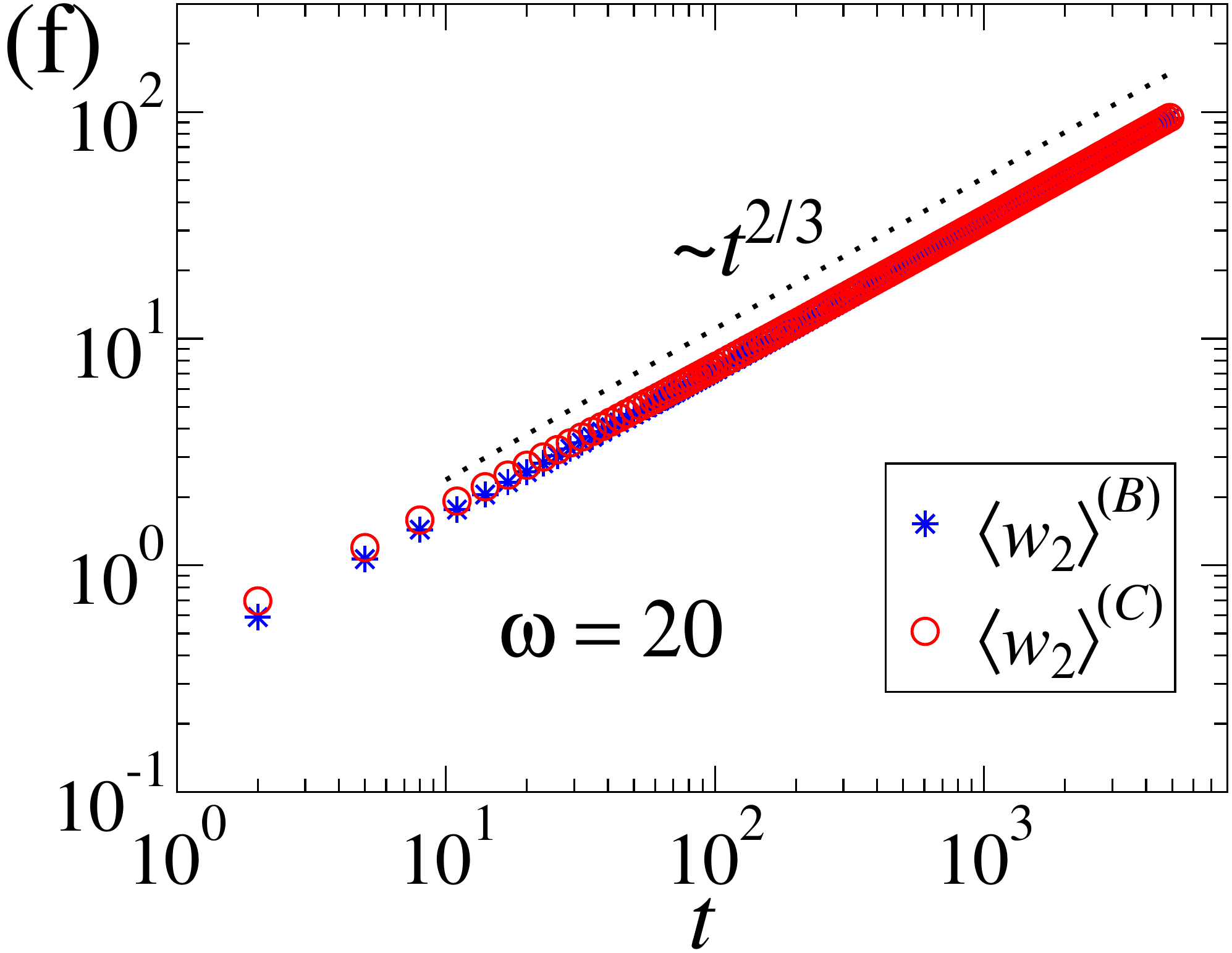}
	\caption{Temporal evolution of the skewness $S$ (top), kurtosis $K$ (middle) and squared roughness $\expct{w_2}$ (bottom panels), for the 1D RSOS model on substrates expanding at rates $\omega=1$ (left) and $\omega=20$ (right panels). The dashed horizontal lines represent the ratios for the TW-GUE distribution, while the dotted lines have the indicated slopes.}
	\label{fig6}
\end{figure}

Figures \ref{fig6}(a) and \ref{fig6}(c) present the temporal variation of $-S$ and $K$, respectively, for the RSOS model simulated with an expansion rate $\omega=1$, comparing results for the three methods. Similarly to what one has observed in the previous section for the flat case, the deviations from the expected ratios, at a given time, are larger in $S^{(A)}$, followed by $S^{(B)}$ and smaller in $S^{(C)}$. This effect is much more drastic in the kurtosis, where $K^{(C)}$ agrees with $K_{GUE}$ already for $t \gtrsim 100$, whereas $K^{(A)}$ and $K_{\kappa}^{(B)}$ have very strong deviations. Milder corrections are observed in $K_{m}^{(B)}$, but it deviates more than $K^{(C)}$. Analogous behaviors are found for the BD and Etching models, confirming that method $(C)$ is indeed very superior than the others to reveal the asymptotic HDs, also for curved KPZ interfaces.

It is important to notice that by increasing $\omega$ the corrections observed for $\omega=1$ decrease. This is demonstrated in Figs. \ref{fig6}(b) and \ref{fig6}(d), which respectively show $-S$ and $K$ versus time for the RSOS model with $\omega=20$. In this case, the skewness calculated from the three methods are very similar. Moreover, one sees that $K^{(C)} \approx K_{m}^{(B)}$ and $K^{(A)} \approx K_{\kappa}^{(B)}$. Although these latter kurtosis are still presenting a slower convergence than the former ones, their deviation from $K_{GUE}$ is much smaller than that for $\omega=1$. Interestingly, while large expansion rates decrease the corrections in the ratios estimated with approaches $(A)$ and $(B)$, the convergence of $S^{(C)}$ and $K^{(C)}$ becomes slightly slower as $\omega$ increases.

I end this section by remarking that the superiority of the ``1-pt statistics'' is not limited to the ratios; it yields also much better results for the roughness in the case of expanding substrates. In fact, a comparison of curves of $\expct{w_{2}}^{(x)}$ versus time, with $x=B$ and $C$, for small $\omega$'s [see Fig. \ref{fig6}(e) for the RSOS model with $\omega=1$] reveals that while $\expct{w_{2}}^{(C)}$ increases consistently with $t^{2\beta}$ already at short times, severe corrections are found in $\expct{w_{2}}^{(B)}=m_2^{(B)}=\kappa_{2}^{(B)}$. For large $\omega$'s, however, the fast increase of the substrate size washes out the corrections and both approaches yield very similar results for the roughness scaling [see Fig. \ref{fig6}(f)].

\section{Discussion}
\label{secDisc}

To understand the origins of the differences in the ratios, as observed above, it is important to analyze the three averaging procedures in more detail. 

In Sec. \ref{secModels} it was demonstrated that $K_{\kappa}^{(B)}=K_m^{(B)} - 3 R_2$ (see Eqs. \ref{eqKAKB} and \ref{eqCoefVar}), where $R_2 = \expct{w_2^2}_c/\expct{w_2}^2$ is the ratio between the second and the (squared) first cumulant of the width distribution, $P_{w_2}(w_2)$. As demonstrated in Ref. \cite{Ismael16b}, these cumulants follow the general scaling
\begin{equation}
\expct{w_2^n}_c \simeq A_n L^{2 n \alpha} f_n\left(t/L^z \right),
\label{eqCumulW2}
\end{equation}
with $f_n(x) = 1$ in the SSR (when $x \gg 1$) and $f_n(x) \sim x^{\gamma_n}$ in the GR (i.e., for $x \ll 1$), where $\gamma_n = 2 n \beta + \frac{(n-1)d}{z}$ \cite{Ismael16b}. This yields $K_{\kappa}^{(B)} \simeq K_m^{(B)} - \left(t^{1/z}/L\right)^d c$ in the GR, where $c$ is a constant, such that asymptotically (for $L \rightarrow \infty$) we shall have $K_{\kappa}^{(B)}=K_m^{(B)}$ in this regime, for both fixed size and expanding substrates, confirming the numerical results above. In the SSR, on the other hand, these kurtoses are indeed different, once $K_{\kappa}^{(B)}=K_m^{(B)} - 3 A_2/A_1^{2}$. For 1D KPZ interfaces, the SSR distributions $P_{w_2}(w_2)$ are exactly known for window boundary conditions (WBC) \cite{antal}, and periodic boundary conditions (PBC) \cite{racz1}, having $R_2 = A_2/A_1^{2} = 2/5$ in the last case. Since the SSR HD about the mean is Gaussian for PBC \cite{barabasi}, meaning that $S^{(B)} = K_m^{(B)} = 0$, one obtains $K_{\kappa}^{(B)}=-6/5$, as indeed observed in Sec. \ref{subsecSSR}. For the 2D KPZ class there is no exact result for these quantities and it seems that numerical estimates of $R_2$ are only available for WBC \cite{fabioLRDs,Ismael16b}. For the 2D RSOS model with PBC, the amplitude $A_1 = 0.1226(1)$ was reported in Ref. \cite{Pagnani} (where it was denoted as $A_2$) and I estimate $A_2 = 0.0020(2)$ here, by extrapolating the saturated values of $\expct{w_2^2}_c/L^{4\alpha}$ to $L \rightarrow \infty$, assuming that $\alpha = 0.3869$ \cite{Pagnani}. This yields $R_2 = A_2/A_1^2 = 0.13(1)$, which agrees with $(K_m^{(B)}-K_{\kappa}^{(B)})/3$ if one uses $K_m^{(B)} = 0.14$ and $K_{\kappa}^{(B)} = -0.22$, as estimated above.

To compare the results from methods $(A)$ and $(B)$, I start noticing that $S^{(B)} = \dfrac{\kappa_3}{\expct{w_2}^{{3}/{2}}} = \dfrac{\kappa_3}{\expct{w_2^{{3}/{2}}}} \left(1 + R_{\frac{3}{2}}\right)$ and $K_{\kappa}^{(B)} =\dfrac{\kappa_4}{\expct{w_2}^{2}} = \dfrac{\kappa_4}{\expct{w_2^{2}}} (1 + R_2)$, with $R_n$ given in Eq. \ref{eqCoefVar}. Note also that, if $a$ and $b\neq 0$ are two fluctuating variables, from a Taylor expansion around $\left< a \right>$ and $\left< b \right>$ one has that $\left< \dfrac{a}{b} \right> = \dfrac{\expct{a}}{\expct{b}} + \sum\limits_{j=1}^{\infty} (-1)^j \left( \dfrac{\expct{a}\expct{(\tilde{b})^j}}{\expct{b}^{j+1}} + \dfrac{\expct{\tilde{a} (\tilde{b})^j}}{\expct{b}^{j+1}}  \right)$, where $\tilde{x} \equiv x - \expct{x}$, so that $\expct{(\tilde{b})^j}$ is the $j$th central moment of $b$ \cite{Rice}. This means that 
\begin{equation}
 S^{(A)} = \left<\frac{(\kappa_3)_i}{w_2^{3/2}}\right> = \frac{\kappa_3}{\expct{w_2^{3/2}}} + G_3 = \frac{S^{(B)}}{1+R_{\frac{3}{2}}} + G_3
\end{equation}
and
\begin{equation}
 K^{(A)} = \left<\frac{(\kappa_4)_i}{w_2^{2}}\right> = \frac{\kappa_4}{\expct{w_2^{2}}} + G_4 = \frac{K_{\kappa}^{(B)}}{1+R_2} + G_4 
\end{equation}
where $G_n = \sum\limits_{j=1}^{\infty} (-1)^j \left( \dfrac{\kappa_n \expct{(\tilde{b}_n)^j}}{\expct{b_n}^{j+1}} + \dfrac{\expct{\tilde{a}_n (\tilde{b}_n)^j}}{\expct{b_n}^{j+1}}  \right)$, with $a_n \equiv (\kappa_n)_i$ and $b_n \equiv w_2^{n/2}$. Therefore, $R_n$ and $G_n$ are responsible for the differences between the ratios from methods $(A)$ and $(B)$. Unfortunately, it is hard to give a full account of the general behavior of $G_n$ (with $t$ and $L$) to proceed further in this comparison. Nevertheless, I remark that for the SSR of 1D KPZ interfaces one has $\kappa_3 = 0$, so that the first term in the summation of $G_3$ is null. Moreover, I have numerically verified that the first mixed moments $\expct{\tilde{a}_3 (\tilde{b}_3)^j}$ also vanish, indicating that $G_3 = 0$ in this case, which explains the saturation values $S^{(A)} = S^{(B)} = 0$ in Fig. \ref{fig5}a. On the other hand, the terms of the sum are non-null for $G_4$, pointing that $G_4 \neq 0$. This is consistent with the numerical findings above, once $K_{\kappa}^{(B)}/(1+R_2) = -6/7 \approx -0.857$ in the SSR of 1D KPZ interfaces, while $K^{(A)} = -0.592(4)$, so that $G_4 \approx 0.265$ in this case.

\begin{figure}[!t]
	\includegraphics[width=4.25cm]{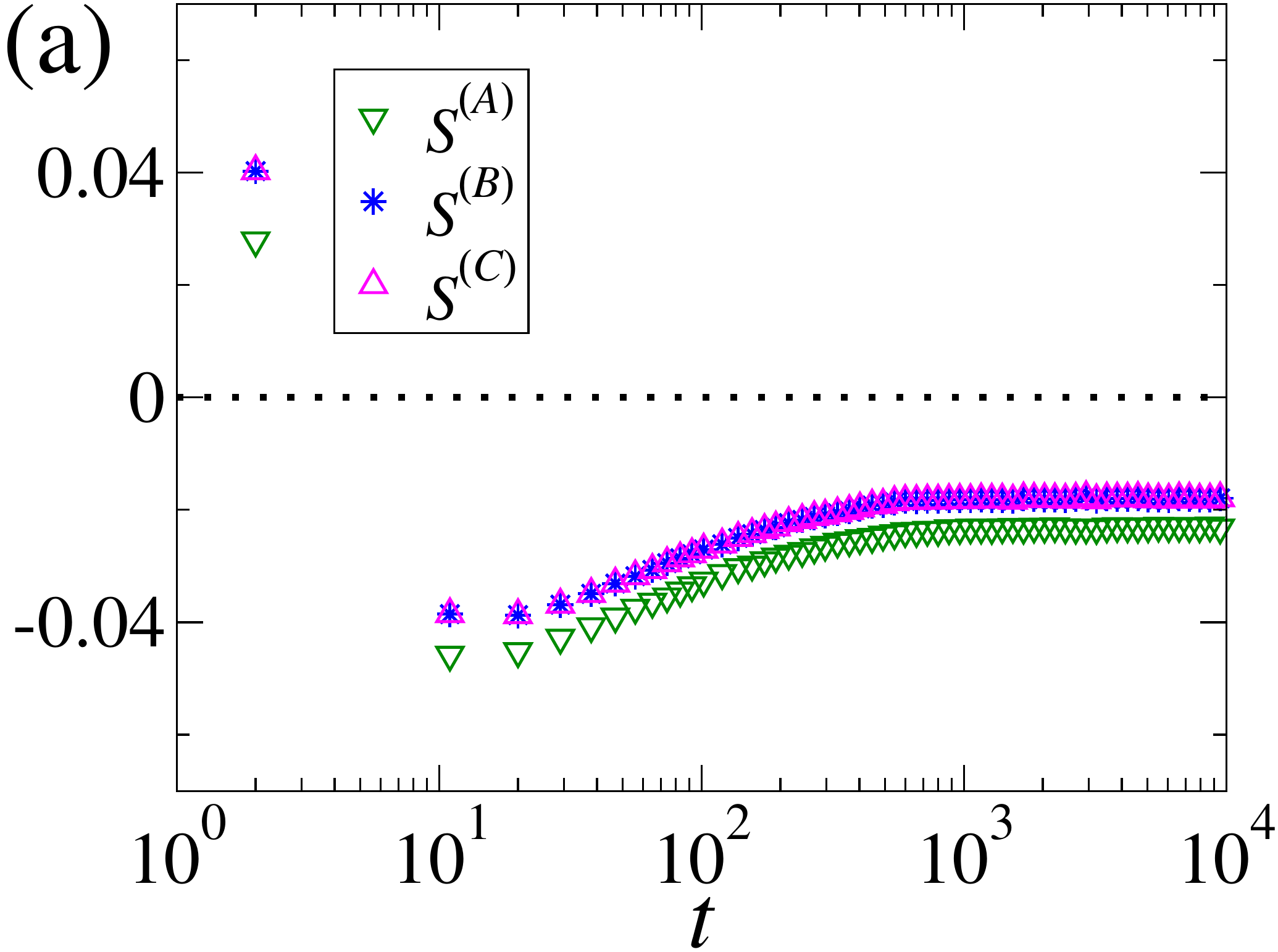}
	\includegraphics[width=4.25cm]{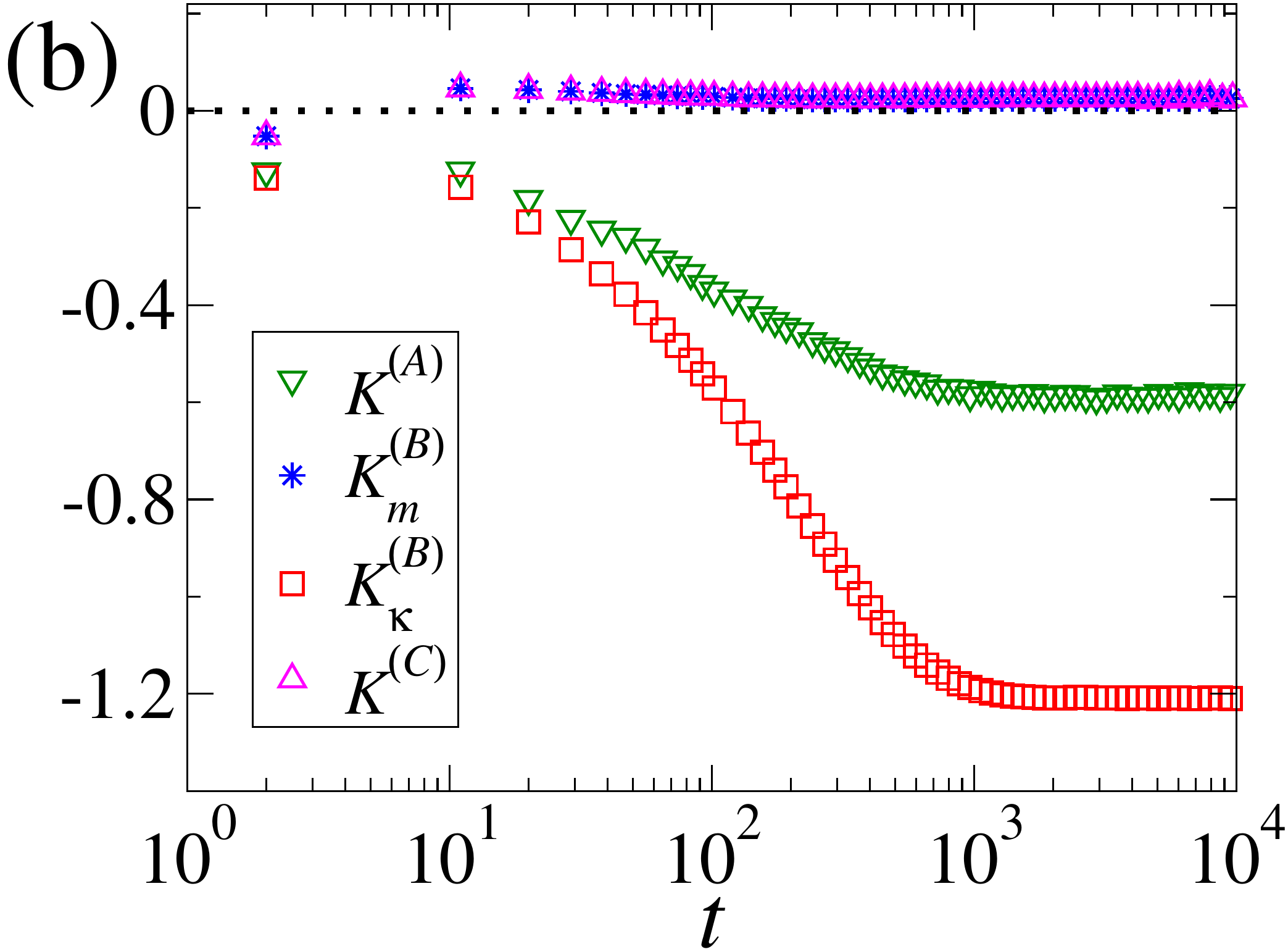}
	\caption{Temporal evolution of the (a) skewness $S$ and (b) kurtosis $K$, for the Family model on a 1D substrate of size $L=128$. Data for approaches $(A)$, $(B)$ and $(C)$ are shown, as indicated.}
	\label{fig7}
\end{figure}

Let us now focus on a comparison of approaches $(B)$ and $(C)$. Note that the height at a given point $j$ of the interface $i$ can be written as $h_{ij}(t) = \bar{h}_i(t) + H_{ij}(t)$, with $H_{ij}(t)$ giving the fluctuations about the average height $\bar{h}_i(t)$. Hence, in systems where $\bar{h}_i(t)$ is deterministic, fluctuations in $h_{ij}$ are identical to those in $H_{ij}$, meaning that $S^{(C)} = S^{(B)}$ and $K^{(C)} = K_m^{(B)}$. An example of such a system is the Family model, where $\bar{h}_i(t) = t$ for all interface $i=1,\ldots,N$. Curves of $S$ and $K$ versus $t$, for the three methods, are shown in Fig. \ref{fig7} for the 1D Family model deposited on a substrate of fixed size $L=128$ and, in fact, no difference between $S^{(C)}$ and $S^{(B)}$, or $K^{(C)}$ and $K_m^{(B)}$ are observed there. Although EW interfaces have an up-down symmetry, such that their HDs are expected to be symmetric, a slight deviation from $S=0$ is seen in Fig. \ref{fig7}(a), due to finite-size effects. Similarly to the KPZ models, the deviation is larger in $S^{(A)}$, confirming that method $(A)$ is the worse to analyze the HDs. As expected, the kurtoses $K^{(A)}$, $K_m^{(B)}$ and $K_{\kappa}^{(B)}$ for the 1D Family model saturate at the same values found for the 1D KPZ models, once in this dimension both classes have equivalent statistics at the SSR. Notwithstanding, the outcomes from approach $(C)$ are different.

\begin{figure}[!t]
	\includegraphics[width=4.25cm]{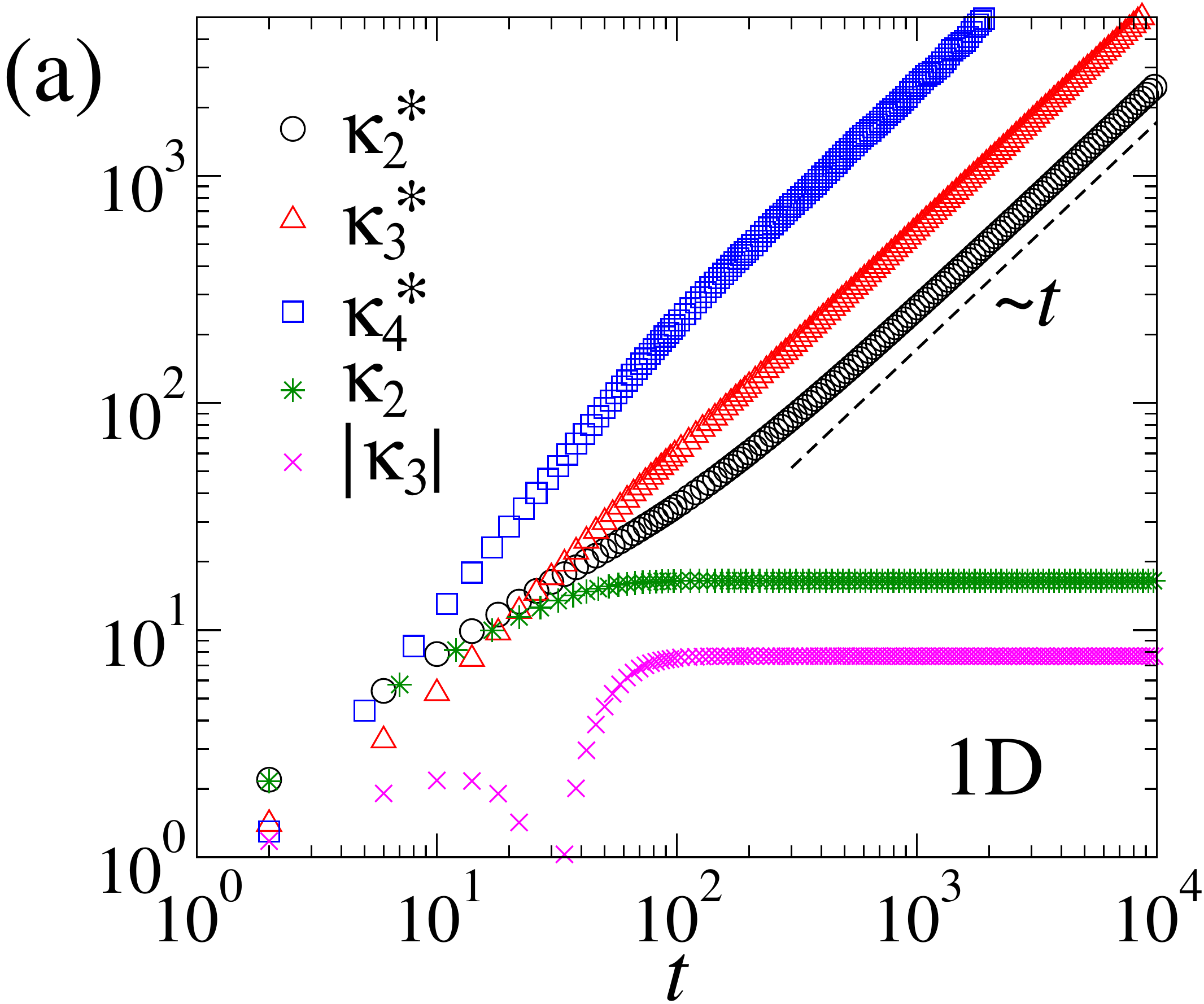}
	\includegraphics[width=4.25cm]{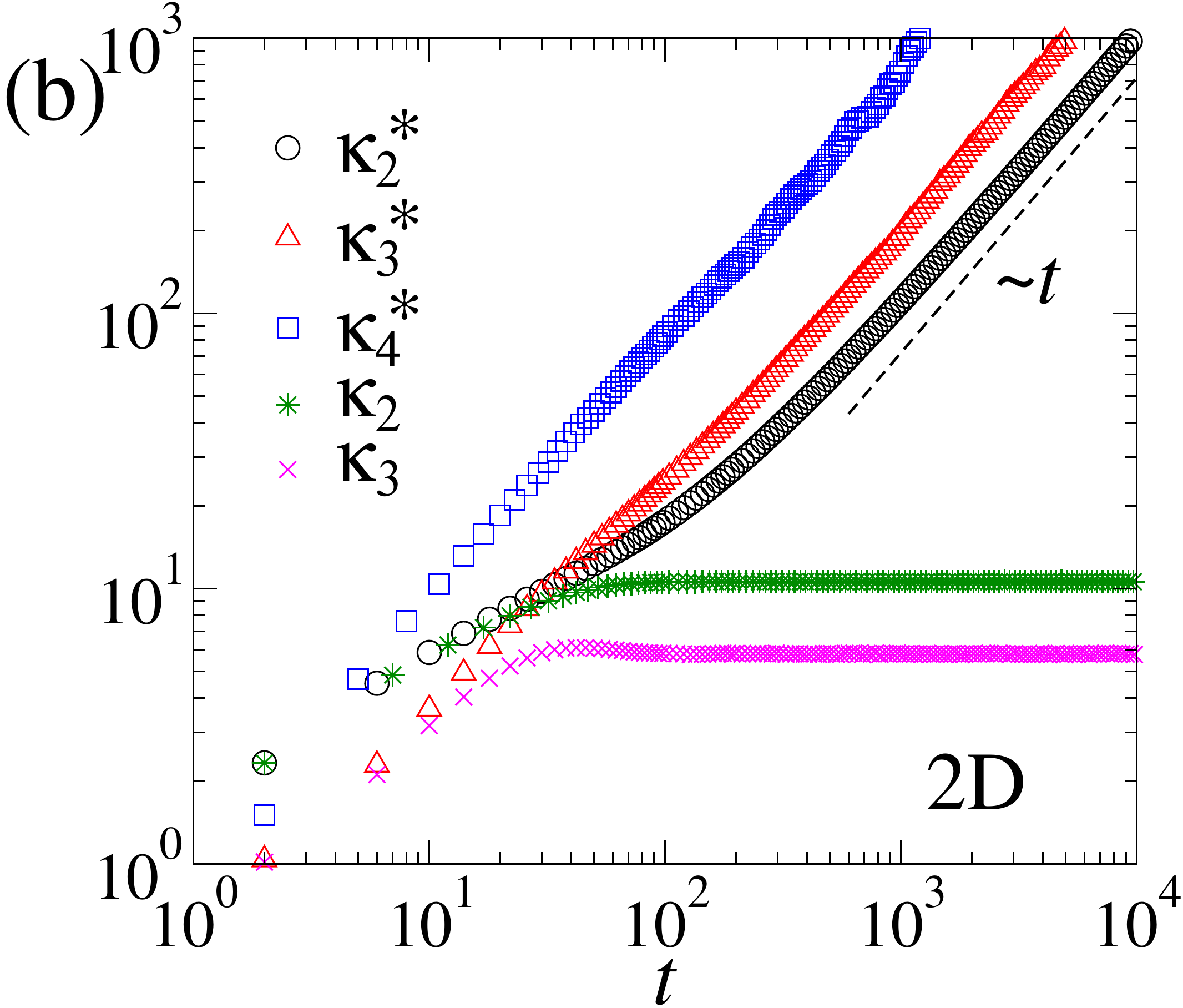}
	\includegraphics[width=4.25cm]{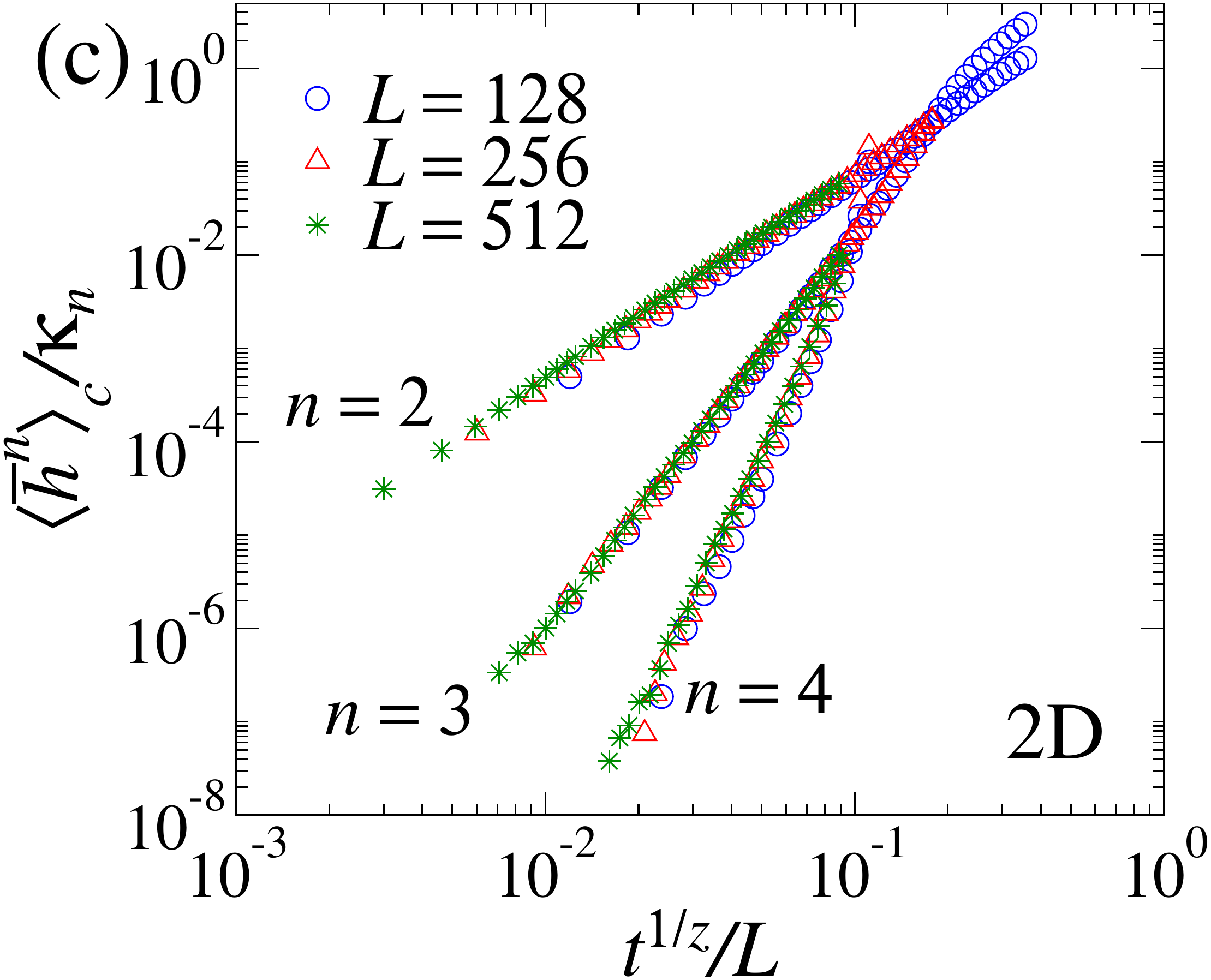}
	\includegraphics[width=4.25cm]{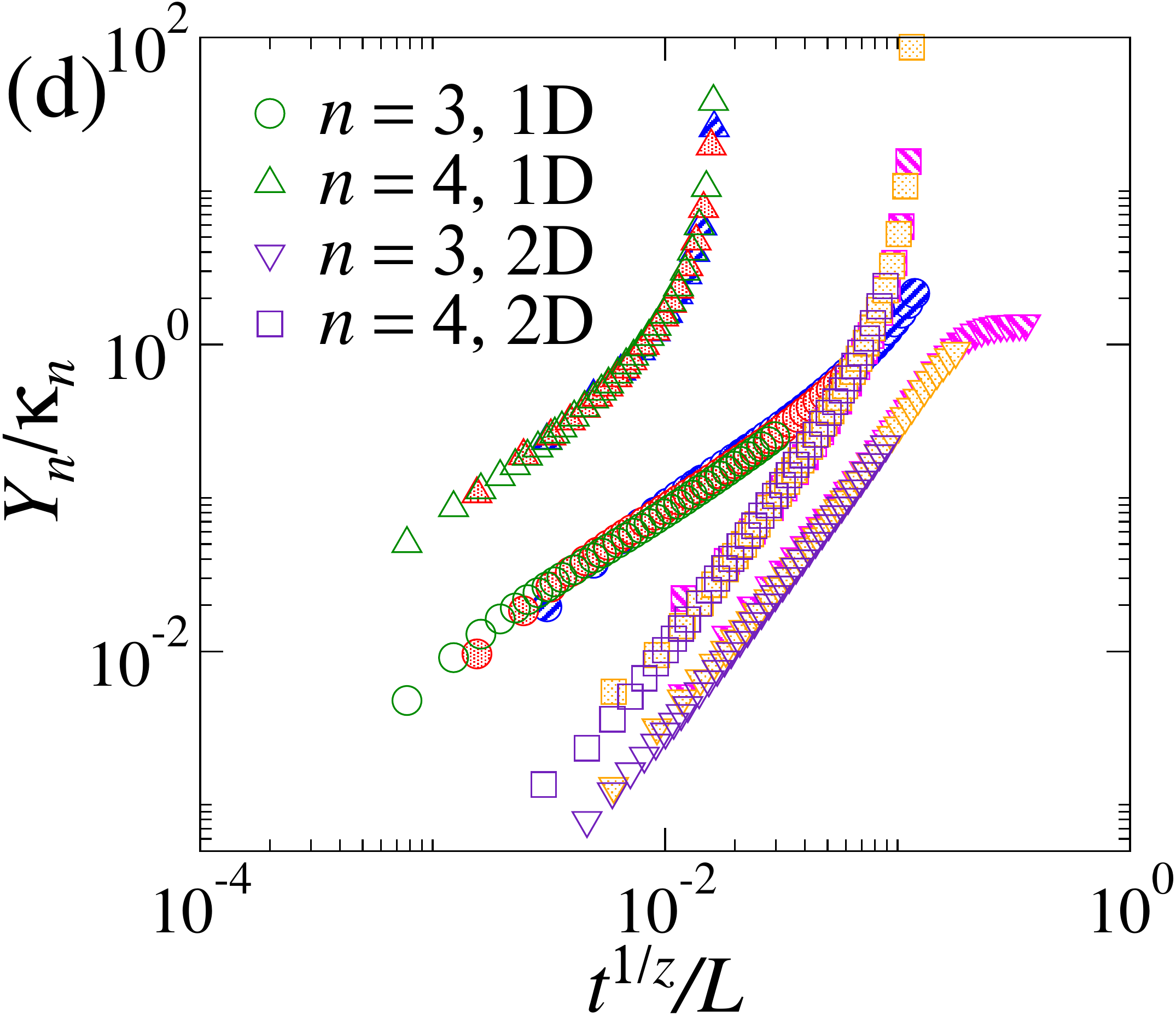}
	\caption{(a)-(b) Temporal behavior of the first cumulants $\kappa_n$ [from method $(B)$] and $\kappa_n^*$ [from method $(C)$], for systems with fluctuating $\bar{h}_i$ on (a) 1D and (b) 2D substrates of lateral size $L=64$. Similar results are found for other $L$'s. The dashed lines have the indicated slope. (c) Cumulant ratio $\expct{\bar{h}^n}_c/\kappa_n$ versus $t^{1/z}/L$, for the 2D case, for several $n$'s and $L$'s, as indicated. (d) Ratio $Y_n/\kappa_n$ against $t^{1/z}/L$, for $n=3,4$ and both 1D (where $z=3/2$) and 2D (where $z = 1.613$ \cite{Pagnani}) substrates, as indicated. The different colors and patterns in the symbols in panel (d) represent data for different sizes in the ranges $L \in [512,2048]$ (1D) and $L \in [128,512]$ (2D). All data are for the Etching model and similar ones are found for the other KPZ models investigated here.}
	\label{fig8}
\end{figure}

This happens because in the KPZ models analyzed here $\bar{h}_i(t)$ is a fluctuating variable, which may assume a different value for each interface $i$. Therefore, the cumulants of $h_{ij}$ --- i.e., those for method $(C)$, denoted by $\kappa_n^*$ --- are given by $\kappa_n^* = \kappa_n + \expct{\bar{h}^n}_c + Y_n$, where $\expct{\bar{h}^n}_c$ is the $n$th cumulant of the distribution of the average height, $P_{av}(\bar{h})$, and $Y_n$ is mostly given by sums of covariances of integer powers of $H_{ij}$ and $\bar{h}_i$. In fact, it is straightforward to demonstrate that $Y_2 = 0$, $Y_3 = 3 cov[H_{ij}^2,\bar{h}_i]$ and $Y_4 = 3 \expct{w_2^2}_c + 4 cov[H_{ij}^3,\bar{h}_i]  + 6 cov[H_{ij}^2,\bar{h}_i^2] - 12 \expct{\bar{h}} cov[H_{ij}^2,\bar{h}_i]$. Thereby, for the ``1-pt squared roughness'', $\kappa_2^*$, one obtains $\kappa_2^* = \expct{w_2} + \expct{\bar{h}^2}_c$. As recently demonstrated in Ref. \cite{tiago21}, $\expct{w_2} \gg \expct{\bar{h}^2}_c$ during the GR, such that $\kappa_2^* \approx \expct{w_2} \sim t^{2\beta}$ there. In the SSR, on the other hand, $\expct{w_2}$ saturates, while $\expct{\bar{h}^2}_c$ keeps increasing forever as $\expct{\bar{h}^2}_c \sim t$, yielding $\kappa_2^* \sim \expct{\bar{h}^2}_c \sim t$ asymptotically. This is indeed confirmed in Figs. \ref{fig8}(a)-(b), which compare curves of $\kappa_n$ and $\kappa_n^*$ against time, for the Etching model on 1D and 2D substrates of fixed size $L=64$, for $n \le 4$. Analogous results are found for the other KPZ models analyzed here.

As is clearly seen in Figs. \ref{fig8}(a)-(b), the higher order cumulants behave similarly to the roughness in the SSR. Namely, while $\kappa_n$ has the expected saturation, the ``1-pt cumulants'' keep increasing as $\kappa_n^* \sim t$. This scaling is indeed expected for the 1D KPZ models, since several works on the large deviation function have demonstrated that $\expct{\bar{h}^n}_c/t$ attains constant values at the SSR in this dimension \cite{DerridaLDF,*DerridaLDF2,*Appert,*LeeLDF,LeeLDF2}. The results here show that the same happens also in the 2D case. Thereby, for these systems one obtains $S^{(C)} = \kappa_3^*/(\kappa_2^*)^{3/2} \sim t^{-1/2}$ and $K^{(C)} = \kappa_4^*/(\kappa_2^*)^{2} \sim t^{-1}$ in the SSR, instead of the saturation observed in approaches $(A)$ and $(B)$. These asymptotic decays are indeed found for all KPZ models analyzed here, in both substrate dimensions [see Figs. \ref{fig1} - \ref{fig3}]. Hence, for such systems, method $(C)$ gives no information on the underlying SSR HDs. This does not mean, however, that is impossible to analyze these HDs by any type of 1-pt fluctuations. For instance, as demonstrated in subsection \ref{subsecAnsatz} just below, by appropriately defining an ansatz for the 1-pt height in the SSR, in the same lines of Eq. \ref{eqAnsatzGR} for the GR, one can access the universal SSR HDs. As an aside, I notice also that by investigating the 1-pt fluctuations of $\Delta h (\vec{x},\Delta t,t_0) \equiv h(\vec{x},t_0+\Delta t)-h(\vec{x},t_0)$ the stationary HD (which is the Baik-Rains \cite{BR} distribution for 1D KPZ systems \cite{Prahofer2000,Imamura}) can be obtained as $t_0 \rightarrow \infty$ and then $\Delta t \rightarrow \infty$ \cite{TakeuchiCross,HealyCross}. 

During the GR, since $\expct{\bar{h}^2}_c/\expct{w_2} \ll 1$ \cite{tiago21}, one may write
\begin{equation}
 S^{(C)} = S^{(B)} \left[1 + \frac{Y_3}{\kappa_3} + \frac{\expct{\bar{h}^3}_c}{\kappa_3} - \frac{3}{2} \frac{\expct{\bar{h}^2}_c}{\kappa_2} + \cdots \right] 
 \label{eqScSb}
\end{equation}
and
\begin{equation}
K^{(C)} = K_{\kappa}^{(B)} \left[1 + \frac{Y_4}{\kappa_4} + \frac{\expct{\bar{h}^4}_c}{\kappa_4} - 2 \frac{\expct{\bar{h}^2}_c}{\kappa_2} + \cdots \right].
 \label{eqKcKb}
\end{equation}
As demonstrated by Lee \& Kim \cite{LeeLDF2}, for 1D KPZ interfaces in the GR $\expct{\bar{h}^n}_c \sim t^{n-2/3}/L^{n-1}$, such that $\expct{\bar{h}^n}_c/\kappa_n \sim [t^{1/z}/L]^{n-1}$, which vanishes as $L \rightarrow \infty$. Although the cumulants of $P_{av}(\bar{h})$ have never been analyzed in literature for 2D interfaces, to the best of my knowledge, from the generality of the scaling behavior it is reasonable to expect that $\expct{\bar{h}^n}_c/\kappa_n \sim [t^{1/z}/L]^{\gamma}$ also in this dimension. This is indeed confirmed in Fig. \ref{fig8}(c), which shows that curves of $\expct{\bar{h}^n}_c/\kappa_n$, for a given $n$ and different $L$'s, collapse onto a single curve when plotted against $[t^{1/z}/L]$. Hence, also in the 2D case, all terms of type $\expct{\bar{h}^n}_c/\kappa_n$, in Eqs. \ref{eqScSb} and \ref{eqKcKb}, vanish asymptotically. Figure \ref{fig8}(d) shows $Y_n/\kappa_n$ versus $[t^{1/z}/L]$, for $n=3,4$ and 1D and 2D systems. Once again, the data for different sizes (for a given $n$ and dimension) collapse, demonstrating that $Y_n/\kappa_n \sim  [t^{1/z}/L]^{\delta}$. So, these terms also become irrelevant in Eqs. \ref{eqScSb} and \ref{eqKcKb} as $L \rightarrow \infty$, confirming that asymptotically one indeed has $S^{(C)} = S^{(B)} = S^{(A)}$ and $K^{(C)} = K_{\kappa}^{(B)}  = K_{m}^{(B)} = K_{\kappa}^{(A)}$ in the GR, as indicated by the results from the previous sections.

It is worth noticing that, in opposition to methods $(A)$ and $(B)$, in case $(C)$ the HDs' cumulants are calculated without using the average height of each interface, so that $S^{(C)}$ and $K^{(C)}$ are not directly affected by the finite-size and -time corrections in $\bar{h}_i$. Moreover, the process of calculating the raw moments for all interfaces together, as done in $(C)$, is analogous to join the $N$ samples to form a single interface of lateral size $N^{\frac{1}{d}} L$, which is expected to have much weaker corrections than those of size $L$ when $N$ is large. In fact, although each interface may have a different $\bar{h}_i$, which might introduce corrections in the composed system of size $N^{\frac{1}{d}} L$, the fluctuations in $\bar{h}_i$ are very small when compared with those in $H_{ij}$ in the GR. This explains the superiority of approach $(C)$ for estimating $S$ and $K$ for the GR HDs.

\subsection{``KPZ ansatz'' for the steady state HDs}
\label{subsecAnsatz}

From the discussion just above, and bearing in mind the ``KPZ ansatz'' in Eq. \ref{eqAnsatzGR} for the 1-pt height at the GR, it is interesting to devise a similar ansatz for the SSR. To do this, let us start recalling that only two non-universal parameters are need to fix the KPZ scaling: $\lambda$ (the non-linear coefficient of the KPZ equation \cite{KPZ}) and $A$ (the amplitude of the height-difference correlation function $C_h = \expct{[h(\vec{x} + \vec{r},t) - h(\vec{x},t)]^2} \simeq A |\vec{r}|^{2\alpha}$) \cite{Krug1992,barabasi,healy95}. In fact, from a dimensional analysis one obtains $\expct{w_2} \sim A L^{2\alpha}$ in the SSR, indicating that
\begin{equation}
 h_{ij} = \bar{h}_i + s_{\lambda} A^{\frac{1}{2}} L^{\alpha} \zeta + \cdots,
 \label{eqAnsatzSSR}
\end{equation}
where $s_{\lambda}=\pm 1$ gives the signal of $\lambda$ for a given KPZ system, $\zeta$ is a random variable expected to be given by universal HDs and the dots indicate the existence of corrections. 

\begin{figure}[!t]
	\includegraphics[width=4.25cm]{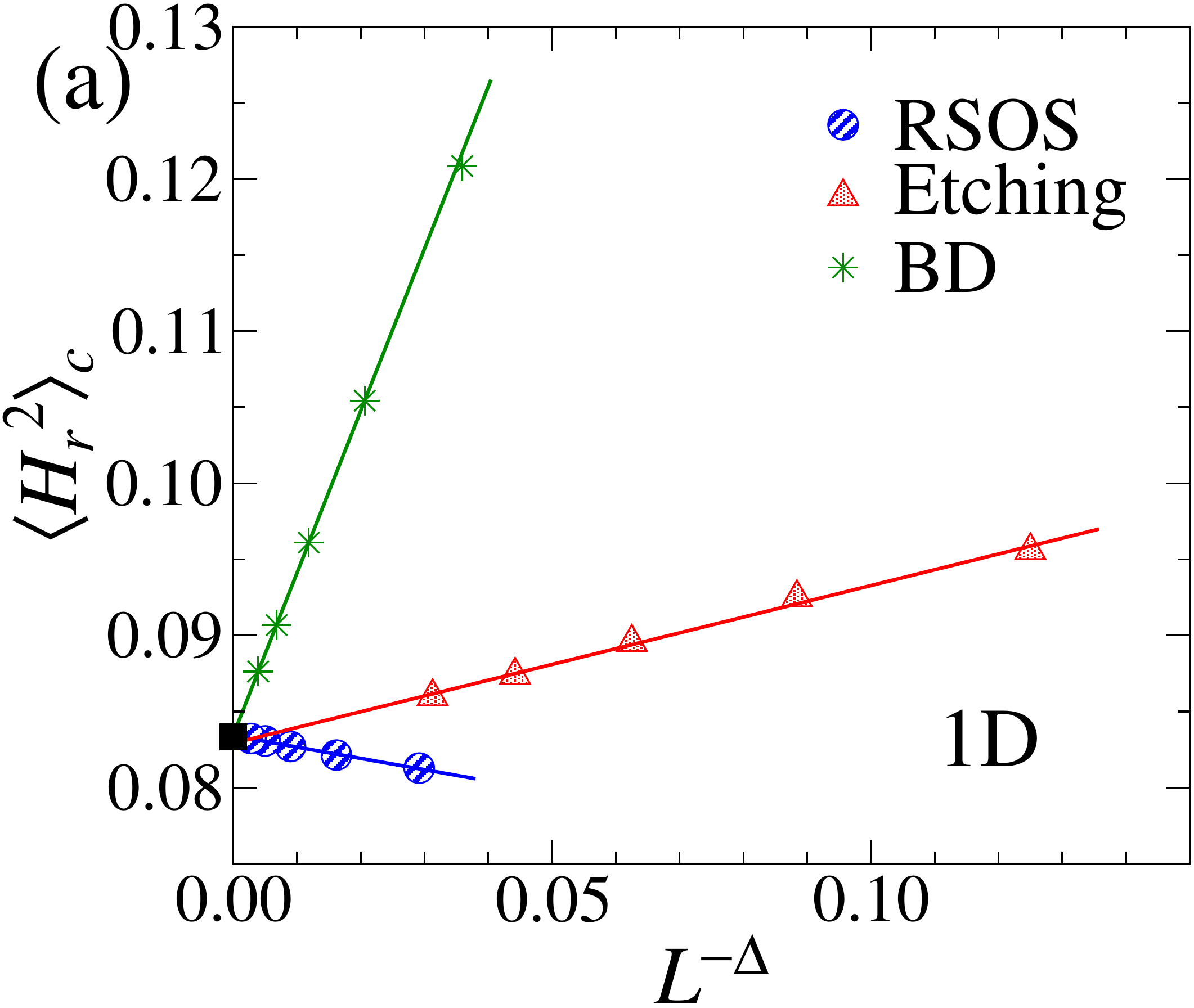}
	\includegraphics[width=4.25cm]{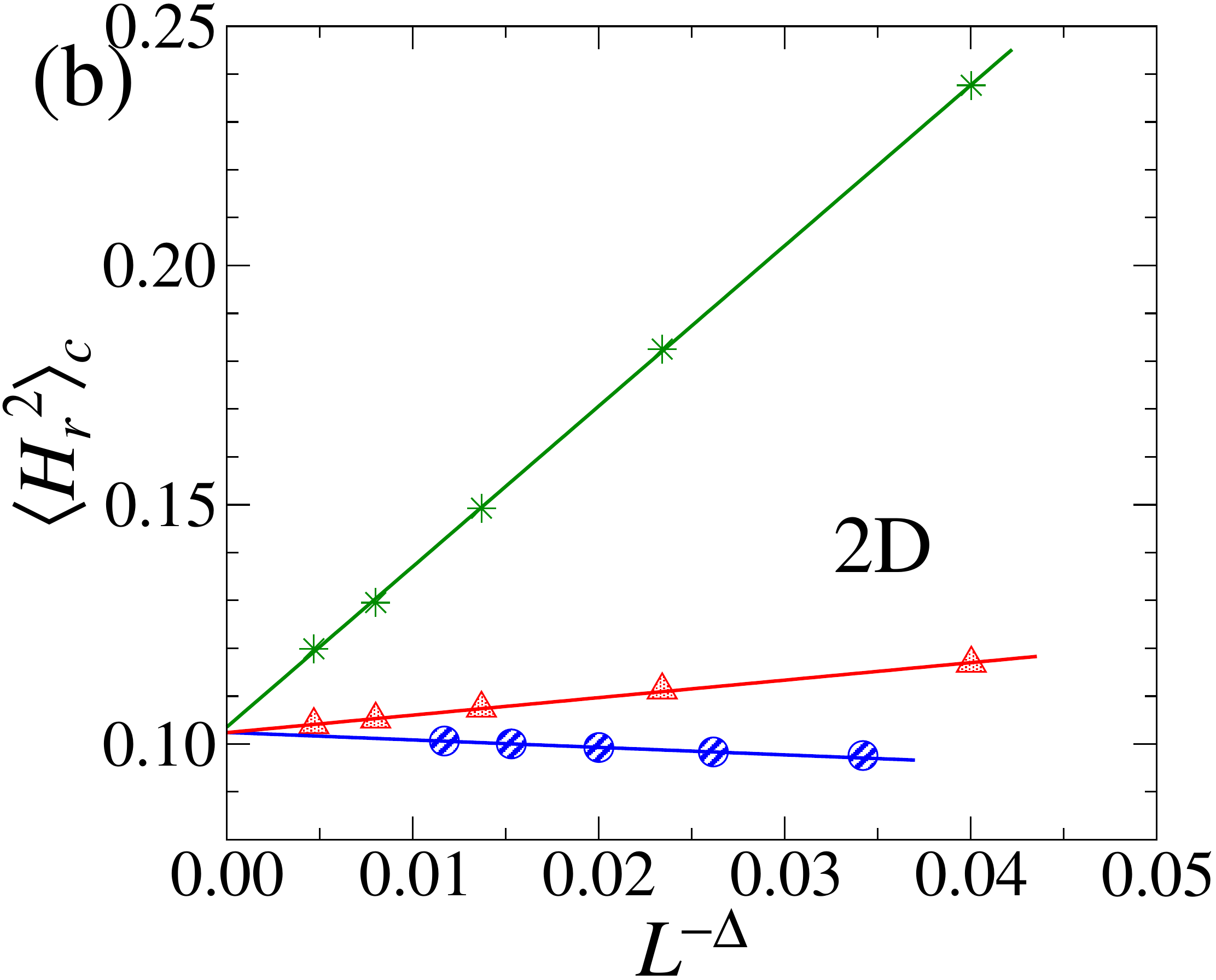}
	\caption{Variances $\expct{H_r^2}_c$ of the 1-pt SSR height fluctuations about the mean versus $L^{-\Delta}$ for the KPZ models on (a) 1D and (b) 2D substrates. The solid lines are linear fits used in extrapolations. The exponents that best linearize the data in each case are $\Delta=\alpha$ [$0.8$] for the RSOS, $\Delta=2\alpha$ [$0.5$] for the Etching and $\Delta=2\alpha$ [$0.8$] for the BD model in the 2D [1D] case. The black dot in panel (a) indicates the exact result $\expct{\zeta^2}_c = 1/12$. The values of $L^{-\Delta}$ for the RSOS model in panel (b) were divided by 10 to appear in the same interval of the rest.}
	\label{fig9}
\end{figure}

As discussed above, in the GR the fluctuations in $\bar{h}_i$ are negligible when compared with those in $H_{ij}$ [given by $(\Gamma t)^{\beta} \chi$ in Eq. \ref{eqAnsatzGR}], justifying the replacement of $\bar{h}_i$ by its asymptotic (and deterministic) behavior $\bar{h}_i \rightarrow v_{\infty} t$ in the ansatz \ref{eqAnsatzGR}. The same thing can not be done in the SSR, once the fluctuations in the rhs of Eq. \ref{eqAnsatzSSR} are dominated by $\bar{h}_i$ in this regime. Notwithstanding, one may uncover the universality of the HD $P(\zeta)$ by analyzing the rescaled relative height 
\begin{equation}
 H_r \equiv \frac{h_{ij} - \bar{h}_i}{s_{\lambda} A^{1/2} L^{\alpha}},
 \label{eqAnsatzSSR2}
\end{equation}
such that $H_r \rightarrow \zeta$ as $L \rightarrow \infty$. 
Since $P(\zeta)$ gives the fluctuations about the mean, it is asymptotically Gaussian-distributed for 1D KPZ and EW systems with PBC, meaning that its higher order cumulants vanish, i.e., $\expct{\zeta^n}_c = 0$ for $n>2$. Moreover, the variance is $\expct{\zeta^2}_c = 1/12$, once the saturated squared width is exactly know to behave asymptotically as $\expct{w_2} = A L / 12$ \cite{Krug1992,KrugAdv,barabasi}. This is indeed verified in Fig. \ref{fig9}(a), where numerical estimates of $\expct{H_r^2}_c$ are extrapolated to $L \rightarrow \infty$ and, for all models analyzed here, attain values very close to $1/12$. In such plots the parameters $A_{RSOS}=0.83$ and $A_{BD}=2.7$ were used, as obtained in Ref. \cite{tiago12a}. Although there is no estimate of $A$ for the 1D Etching model in the literature, there exist several numerical evidence that its scaling amplitudes are equal to those for the BD model. In fact, by using $A_{Etch} = 2.7$ in Fig. \ref{fig9}(a), one obtains $\expct{\zeta^2}_c$ for the Etching case in striking agreement with the other models and with the exact value $1/12$.

For the 2D KPZ models, the values $A_{RSOS} = 1.22$ and $A_{Etch} = 3.629$ were explicitly estimated in Ref. \cite{Ismael14}. Moreover, from the parameters reported in Ref. \cite{Alves14BD} for the BD model, it is simple to calculate $A_{BD} \approx 3.626$, which provides additional evidence that $A_{Etch}=A_{BD}$. Figure \ref{fig9}(b) shows extrapolations of the variances $\expct{H_r^2}_c$ to the $L \rightarrow \infty$ limit, which return very consistent results, being $\expct{\zeta^2}_c = 0.1024$, $0.1023$ and $0.1035$ for the RSOS, Etching and BD models, respectively. Therefore, this yields $\expct{\zeta^2}_c = 0.1027(5)$ for the 2D KPZ height fluctuations about the mean in the SSR. Joining this result with the values obtained above for $S^{(B)}$ and $K_m^{(B)}$ one finds also estimates for $\expct{\zeta^3}_c$ and $\expct{\zeta^4}_c$, which are depicted in Tab. \ref{tab1}, along with the exact results for the 1D case. I have verified that extrapolations of $\expct{H_r^n}_c$ to $L \rightarrow \infty$, for $n=3,4$, return cumulants in agreement with those in Tab. \ref{tab1}, as expected.

I notice that an additive, stochastic correction, $\eta$, has been widely observed in Eq. \ref{eqAnsatzGR} (see, e.g., Refs. \cite{Takeuchi2010,Takeuchi2011,Alves11,tiago12a,tiago13,Ismael14}). Hence, a similar correction might be expected also in Eq. \ref{eqAnsatzSSR}, such that $h_{ij} = \bar{h}_i + s_{\lambda} A^{1/2} L^{\alpha} \zeta + \eta + \cdots$, with $\expct{\bar{\eta}} = 0$. In fact, this yields
\begin{equation}
\expct{H_r^2}_c = \expct{\zeta^2}_c +  s_{\lambda} \frac{2 \expct{\eta \zeta}}{A^{1/2}} L^{-\alpha} + \frac{\expct{\eta^2}_c}{A}  L^{-2\alpha} + \cdots 
\label{eqCorrections}
\end{equation}
Since $s_{\lambda} = -1$ for the RSOS model, assuming that $\expct{\eta \zeta}>0$, this explains why the variances converge from below in this case, while in the other models (where $s_{\lambda}=1$) the convergence is from above (see Fig. \ref{fig9}). Note also that the variances for the Etching in $d=1$ and RSOS in $d=2$ are well linearized with exponents $\Delta = \alpha$, suggesting that $2 \expct{\eta \zeta}/A^{1/2}$ is not negligible in such models. However, for the 2D Etching and BD models this term seems to be negligible, once the dominant correction is $\sim L^{-2\alpha}$. In other cases, the effective correction exponent is $\alpha < \Delta < 2\alpha$, possibly due to a ``competition'' of both correction terms in Eq. \ref{eqCorrections}.

\begin{table}[!b] \centering
\caption{First cumulants of the SSR HDs, from the ansatz \ref{eqAnsatzSSR}, for KPZ systems on 1D and 2D substrates with PBC.} 
\begin{tabularx}{\columnwidth}{p{0.6cm} >{\centering\arraybackslash}X >{\centering\arraybackslash}X >{\centering\arraybackslash}X >{\centering\arraybackslash}X}
 \hline
 \hline
            & $\expct{\zeta}$  & $\expct{\zeta^2}_c$  & $\expct{\zeta^3}_c$  & $\expct{\zeta^4}_c$  \\
  \hline
  1D        & $0$              & $1/12$               & $0$                  & $0$                   \\
  2D        & $0$              & $0.1027$             & $0.0088$             & $0.0015$               \\ 
 \hline
 \hline
\end{tabularx}
\label{tab1}
\end{table}

\section{Summary}
\label{secConc}

I have investigated different procedures for estimating the skewness, $S$, and kurtosis, $K$, of the height fluctuations for a given set of translation-invariant KPZ and EW interfaces, deposited on both fixed-size and enlarging substrates. In general, method $(A)$ --- where $S$ and $K$ are calculated for each interface and then averaged over the $N$ samples --- presents the stronger finite-size corrections. In approach $(B)$ the cumulants, $\kappa_n$, or central moments, $m_n$, of each interface are estimated and then averaged over the $N$ samples, with $S$ and $K$ being calculated only at the end. Hence, $S^{(B)}$ and $K_{m}^{(B)}$ are related to the height fluctuations about the mean height, $\bar{h}_i$, of each interface $i$. Different kurtosis are obtained from this method, with $K_{\kappa}^{(B)} = K_{m}^{(B)} - 3 R_2$, due to the width fluctuations, $P_{w_2}(w_2)$, whose squared variation coefficient is $R_2>0$. Procedure $(C)$ consists in evaluating the raw moments considering all points of the $N$ interfaces together, from which $S$ and $K$ are then calculated. This is equivalent to analyze the height fluctuations at a single point of the interface, so that $S^{(C)}$ and $K^{(C)}$ are related to the ``1-pt'' HDs.

As expected, in systems where $\bar{h}_i$ evolves deterministically in time, the ``1-pt fluctuations'' become identical to the fluctuations about the mean, so that $S^{(C)}(t) = S^{(B)}(t)$ and $K^{(C)}(t) = K_{m}^{(B)}(t)$. On the other hand, these equalities are broken when $\bar{h}_i$ is a fluctuation variable and, then, one obtains $S^{(A)}(t) \neq S^{(B)}(t) \neq S^{(C)}(t)$ and $K^{(A)}(t) \neq K_m^{(B)}(t) \neq K_{\kappa}^{(B)}(t) \neq K^{(C)}(t)$. During the GR, the signature of the universal HDs in finite-$L$ systems is a maximum in the curves of $S \times t$ and $K \times t$, whose values converge to the asymptotic ones as $L \rightarrow \infty$. It turns out that for procedures $(A)$ and $(B)$ such maxima may only appear for very large $L$'s, depending on the model, and may display a slow convergence. Remarkably, method $(C)$ presents very weak corrections in these quantities, with curves of $S^{(C)}(t)$ and $K^{(C)}(t)$ having maxima or approximated plateaus very close to the asymptotic values already for small $L$'s and $t$'s. In addition, the roughness may also have weaker corrections to the scaling, $\expct{w_2} \sim t^{2\beta}$, when estimated from procedure $(C)$ in expanding (or curved) systems. These findings might be particularly important in experimental studies of HDs, where one usually deals with small interfaces and short deposition times, provided that different samples are statistically equivalent to be reliably analyzed via method $(C)$.
Furthermore, since the maximum in the $S^{(C)}(t)$ and $K^{(C)}(t)$ curves can be estimated with high accuracy for small $L$'s in numerical works, their extrapolation to $L \rightarrow \infty$ is a very good route to access the asymptotic values of these ratios for the GR HDs. In fact, this might be much less computationally demanding than the strategy adopted in several previous works, considering very large $L$'s and times, and extrapolating the curves of $S(t)$ and $K(t)$ for the $t \rightarrow \infty$ limit.

\begin{table}[!t] \centering
\caption{Absolute value of the skewness, $|S|$, and kurtosis, $K$, from methods $(A)$ and $(B)$, for the KPZ HDs in the steady state regime for 1D and 2D substrates with PBC.}
\begin{tabularx}{\columnwidth}{p{0.6cm} >{\centering\arraybackslash}X >{\centering\arraybackslash}X >{\centering\arraybackslash}X >{\centering\arraybackslash}X >{\centering\arraybackslash}X}
 \hline
 \hline
            & $|S^{(A)}|$  & $|S^{(B)}|$  & $K^{(A)}$  & $K_m^{(B)}$  & $K_{\kappa}^{(B)}$  \\
  \hline
  1D      & $0$          & $0$          & $-0.592$   & $0$          & $-6/5$              \\
  2D      & $0.232$      & $0.266$      & $-0.134$   & $0.14$       & $-0.22$             \\ 
 \hline
 \hline
\end{tabularx}
\label{tab2}
\end{table}

Since in the GR the ratios from the ``1-interface'' [method $(A)$], ``multi-interface'' [$(B)$] and ``1-pt statistics'' [$(C)$] converge to the same asymptotic values, they are all related to a single HD. In the SSR, on the other hand, different asymptotic results can be obtained from these statistics. For example, one finds $S^{(C)} \sim t^{-1/2}$ and $K^{(C)} \sim t^{-1}$ whenever $\bar{h}_i$ is a fluctuating variable, so that method $(C)$ does not provide information on the universality of the SSR HDs for these KPZ systems. Using the ansatz introduced in Eq. \ref{eqAnsatzSSR} for the 1-pt height in the SSR, however, it is possible to access the universal cumulants and their ratios for the underlying SSR HD. This ``1-pt'' approach via the ansatz is equivalent to method $(B)$ for the moments, once they both give the fluctuations about $\bar{h}_i$. A different, but universal, set of asymptotic values for $S$ and $K$ are obtained from approach $(A)$. 
All the ratios obtained here for the SSR are summarized in Tab. \ref{tab2} for the 1D and 2D KPZ class (mostly estimated for the RSOS model), which might be very useful for later reference. In fact, while all previous works have focused on a single method/quantity to investigate the universality of the SSR HDs, this variety of values makes it clear that, if one wants to use these HDs to demonstrate that a given system belongs to the KPZ class, the most reliable way to do this is by calculating all these ratios. When doing this, it is important to have in mind that fluctuations in the SSR may depend on the boundary conditions and all results presented here are for the periodic case.

\acknowledgments

The author acknowledges financial support from the CNPq and Fapemig (Brazilian agencies), and thanks I.S.S. Carrasco and R.A.L. Almeida for motivating discussions.

\bibliography{bibHDsAverages}

\end{document}